\documentclass[useAMS,usenatbib]{mn2e}
\usepackage{graphicx}
\usepackage{amsmath}
\usepackage{amsfonts}

\DeclareGraphicsExtensions{.pdf, .jpg, .png}

\graphicspath{{./}}

\title[The non-separable GDSAI]{On radial anisotropy limits in stellar systems}
\author[Jeremy A. Barber, Hongsheng Zhao]{Jeremy A. Barber$^{1}$\thanks{E-mail:
jab22@st-andrews.ac.uk (JAB); hz4@st-andrews.ac.uk (HZ)}, Hongsheng Zhao$^{1}$\\
$^{1}$Scottish Universities Physics Alliance, University of St Andrews, North Haugh, St Andrews, Fife, KY16 9SS, UK
}
\begin{document}

\date{Accepted ----. Received ----:}

\pagerange{\pageref{firstpage}--\pageref{lastpage}} \pubyear{2014}
\maketitle

\label{firstpage}

\begin{abstract}
Following earlier authors, we re-examine constraints on the radial velocity anisotropy of generic stellar systems using arguments for phase space density positivity, stability, and separability. It is known that although the majority of commonly used systems have an maximum anisotropy of less than half of the logarithmic density slope \emph{i.e.} $\beta < \gamma/2$, there are exceptions for separable models with large central anisotropy. Here we present a new exceptional case with above-threshold anisotropy locally but with an isotropic center nevertheless. These models are non-separable and we maintain positivity. Our analysis suggests that regions of above-threshold anisotropy are more related to regions of possible secular instability, which might be observed in self-consistent galaxies in a short-lived phase.  
\end{abstract}


\section{Introduction}

Real and simulated stellar systems are often anisotropic as the lack of two-body collisions allows anisotropy from the initial configuration of phase space to persist in equilibrium. Radial anisotropy is difficult to measure observationally because of the lack of 3D velocity information. This, in turn, widens the uncertainty of our estimates of the mass and gravity of galaxies and black holes using the traditional Jeans equation approach. It is thus desirable to set some limits on this anisotropy from arguments such as positivity, stability, and even separability of the underlying phase space density.  

Usually a particular potential or density profile will be chosen to model a particular system of interest. The most effective and powerful presentation of such a system is the phase-space distribution function (DF) which is connected to observable, real-space quanitites of a system via various integral relations. Because the DF is a probability distribution that describes the phase-space of a system there are some fundamental requirements for a DF that produces a viable system. The most basic constraint is the positivity of the DF over the entire permitted domain of the system as while a system with a positive DF may not be stable, but a system with a negative DF cannot even be created.

The relationship between a density profile and a DF is complicated and is not even one-to-one \citep{Dejonghe1987}. Since the DF describes the full six-dimensional shape of the system there are multiple possible DFs that can produce the same the density profile that only differ through, for example, their anisotropy profiles. Accordingly, it is very important to be able to derive unambiguous analytical expressions for a system of interest so that the positivity can be known precisely.

The main problem here is that the process of finding an expression for the DF of an arbirtrary system is highly non-trivial and can usually not be done analytically. The most reliable method of finding a DF is through Eddington's formula \citep{Eddington1916} which inverts the integral relationship between the density and the DF, however even this is only analytic for a selection of density profiles and parameters.

So in general, while specific models and schemes to produce analytical DFs for a given density do exist, there is a pressing need for simple, fundamental relationships between the quantities of a system that can constrain the positivity of a DF. A way to look at a particular model and know, without having to work through the inversions, whether or not the DF is likely to be positive would be ideal.

One particularly important result was that of \citet{Ciotti1992} who found a simple criteria for the consistency of models using an Osipkov-Merritt anisotropy scheme. This paved the way for a dramatic expansion in the scope of such relations, bringing us to the birth of the result we will be examining. The first major step towards a completely general analytical constraint was made by \citet{Hansen2004} in the form of hard constraints on the conditions in the centre of a dark halo under reasonable assumptions of spherical symmetry, a power law phase-space density \citep{Taylor2001}, and the requirement for physical solutions to the Jeans Equations. They found that any system with an inner density profile $\rho \propto r^{-\gamma}$ would obey $1+\beta\leq\gamma\leq 3$ where $\beta$ is the velocity anisotropy parameter.

This was subsequently improved until the relation could constrain a non-negative DF \citep{An2006} in multi-component systems \citep{Ciotti2009} and Cuddeford models \citep{Cuddeford1991,Ciotti2010a} which contain the Osipkov-Merritt models as a special case. After the discovery that the constraints held even for system outside these model groups \citep{Ciotti2010b} there was an effort made to define exactly how universal such constraints could be. This lead to the significant result of \citet{Ciotti2010} where it was proven that a large class of models obey the relation:

\begin{equation}
\label{eqn:GDSAI}
\gamma\geq 2\beta
\end{equation}

This relationship was termed the Global Density Slope-Anisotropy Inequality (GDSAI) and was shown to be strongly connected to the positiity of the DF in this broad class of multi-component models Cuddeford models as well as in a variety of other anisotropic systems. Specifically, the work of \citet{vanhese2011} showed that obeying the GDSAI is a necesary condition for DF postivity in models where the central anisotropy was $\beta_0\leq0$ but did demonstrate counter-examples for larger anisotropies.

All the systems that had been investigated and had a proven relationship to the GDSAI fall into the category of models with separable augmented density. An augmented density is one that can be described only in terms of a potential as a function of radius and the radius itself. A separable model of this kind can be described thusly:

\begin{equation}
\rho(r)=\rho_{aug}(\psi(r),r)=f(\psi)g(r)\quad0\leq\psi\leq\psi_0
\end{equation}
where we alter the usual notation for the augmented density to avoid later confusion with our dimensionless variables.

Since the GDSAI has been proved for all separable augmented systems with $\beta_0\leq0$ and is understood in such systems with $\beta_0>0$, we will investigate the behaviour of augmented systems which are non-separable. We accomplish this by using mono-energy DFs that produce non-separable density profiles which, whle highly artificial, are also comparatively easy to understand and analyse.

We present a simple spherical model that significantly violates the GDSAI over a range of radii, produces systems with $\beta_0=0$, and has a globally positive DF. The DF is a mono-energy halo that is separable in E and $\text{L}^2$. We suggest that this is evidence that the GDSAI cannot be extended to all non-separable systems and cannot be used to constrain the positivity of their DFs. We instead suggest that, since our DF is not guaranteed to be dynamically stable, system stability is still the principle measure that can confirm whether such non-separable systems can be created and kept in equilibrium.

In \S2 we briefly confirm the inadequacies of a purely Jeans Equation-based approach, \S3 shows our construction of a simple system that does not follow the inequality, \S4 examines the practical implications of the system, \S5 examines the stability of the system, \S6 describes the generalisation of our model, and \S7 concludes.

\section{The inadequacy of a Jeans' equation approach}
\label{sec:jeans}

It is already known that the criteria provided by the Jeans' equations are not as stringent as testing for the positivity of the distribution function. However, there are mathematical difficulties associated with calculating properties of most general DF's which mean the Jeans' equations are still relevant.

We will demonstrate why apparently simple violations of the GDSAI which rely on the Jeans' equations are insufficient to disprove it, as noted in \cite{Ciotti2010}. This is done by creating a density profile from two simple, superimposed models which, together, should apparently break the GDSAI according to the Jeans' equation. We will then show why the model fails to achieve this by being unphysical in a way that the Jeans' equations cannot indicate.

\begin{figure}
\includegraphics[width=84mm]{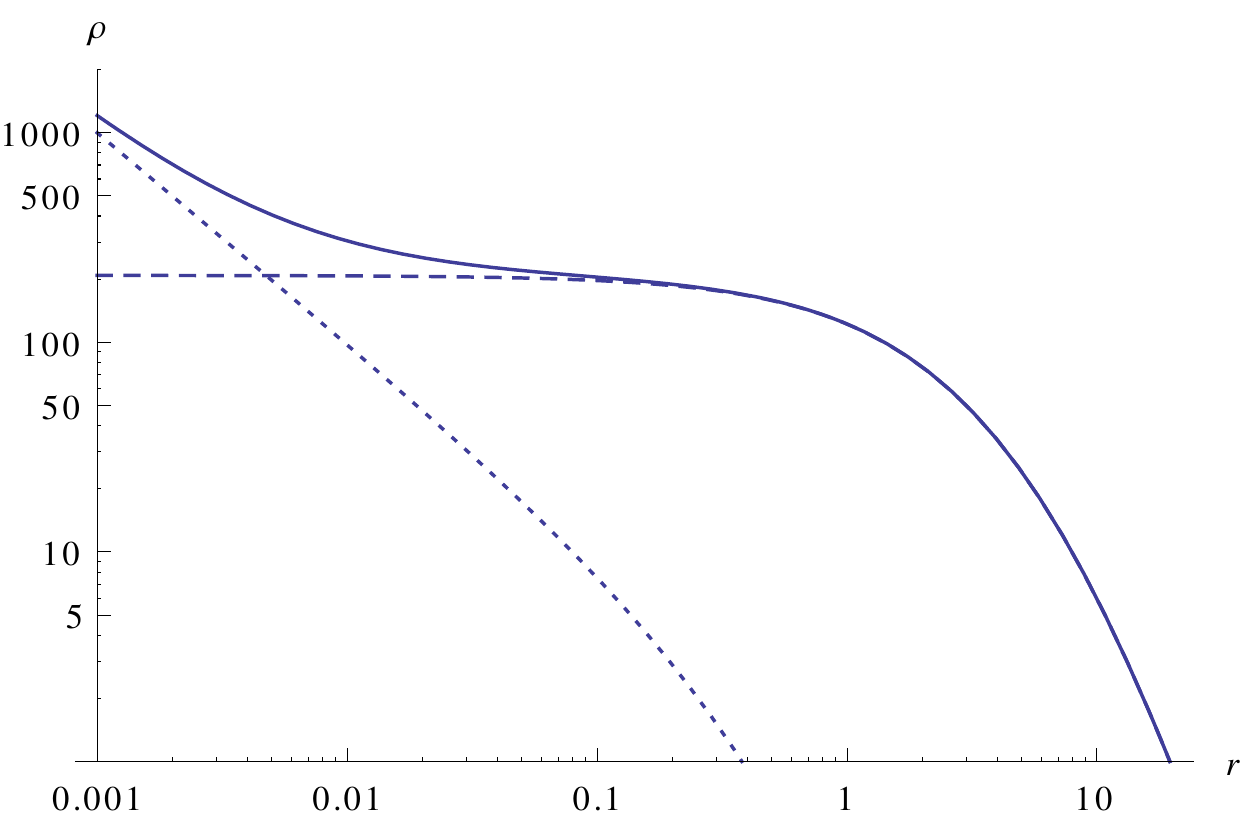}
\caption[Density of Jeans' model for GDSAI violation]{\label{fig:jeansineqden}Density profiles of a simple composite model designed to violate the GDSAI. The dashed and dotted lines are the \mbox{$1/r(1+r)^3$} cored and \mbox{$5\times10^5/(7+r^4)$} cusped subsidiary models while the solid line is the sum of the two profiles.}
\end{figure}

So, the system we construct is created by overlaying a large cusped model from the \citet{Zhao1996} family of models onto a smaller, cored, Hernquist model \citep{Hernquist1990} to create a structure with the density profile shown in Fig. \ref{fig:jeansineqden} described by:

\begin{gather}
\rho=\frac{1}{r(1+r)^3}+\frac{5\times10^5}{(7+r)^4} \\[5pt]
\phi=-\frac{2\pi(7000147+8500042r+1500003r^2}{3r(1+r)(7+r)^2}
\end{gather}

The smaller model has a cusp with $\rho\propto r^{-1}$ in the centre transitioning to $\rho\propto r^{-4}$ while the larger of the two models is cored. This means the constant density core extends past the point where the smaller model has declined to $r^{-4}$. This creates a region in between the $r^{-1}$ cusp and the $r^{-4}$ halo where the Hernquist profile starts to dominate leading to a flattening of the density profile. In this region the density slope is very close to zero, so if we state that our system has an anisotropy of $\beta=1/2$ everywhere then the system will not follow the inequality.

So, we then attempt to solve the Jeans' equation for the system:

\begin{equation}
-\frac{\mathrm{d}(\rho\sigma_r^2)}{\rho\mathrm{d}r}-\frac{2\beta\sigma_r^2}{r}=\frac{\mathrm{d}\phi}{\mathrm{d}r}
\end{equation}

We want to solve this for $\sigma_r^2$ so we can express this as follows assuming constant $\beta$:

\begin{equation}
\rho r^{2\beta}\sigma_r^2=\int^{\infty}_r \rho r^{2\beta}\frac{\mathrm{d}\phi}{\mathrm{d}r}\,\mathrm{d}r
\end{equation}

By definition, we can replace $\frac{\mathrm{d}\phi}{\mathrm{d}r}$ with terms of density instead:

\begin{equation}
\rho r^{2\beta}\sigma_r^2=\int^{\infty}_r \rho r^{2\beta}\frac{G}{r^2}\left[\int_0^r 4\pi r^2\rho\,\mathrm{d}r\right]\,\mathrm{d}r
\end{equation}

A full, rigorous treatment of this integral can be performed, however the density function is sufficiently complex that the analytical result is too large to be worth reproduction here. The result is a collection of hypergeometric series which depend, in part, upon $\beta$.

The problem is that the result for $\sigma_r^2$ is not defined for all values of $\beta$. If $\beta$ is such that $\beta\geq1/2$ then our expression will include instances of evaluating $1/0$ which is undefined. In other words, we cannot evaluate $\sigma_r^2$ for $\beta\geq1/2$ meaning that any attempt to force an anisotropy which violates the GDSAI results in an unphysical solution.

The problem is that it is difficult to predict this failure in advance of solving a specific instance of the Jeans' equation. The separate components of the model are both independently stable, so nothing immediately seems to be wrong with the system that we attempted to create. Rather than working through large numbers of possible models to find combinations of parameters that work, it is easier to work directly with distribution functions.

%
%

For instance, with the benefit of a little \emph{a priori} knowledge, we could have known this system would not be physical. In a system such as this we can assume that the distribution function follows the form \citep{Cuddeford1991}:

\begin{equation}
f{(E,L)}=L^{-2\beta}F(E)\,\textrm{; }F(E)|_{E=\phi}=-\frac{1}{2\pi^2}\frac{\mathrm{d}(r\rho)}{\mathrm{d}\phi}
\end{equation}
where we assumed the form of the function $F(E)$ according to \citet{An2006} assuming that $\beta=1/2$. The problem is that the angular momentum term will always be positive but if we look at the energy term there is one region where $F(E)$ is locally negative.  We can see this more clearly by breaking down the expression for the energy function:

\begin{equation}
\label{eqn:energyfunction}
 F(E)\propto-\frac{\mathrm{d}(r\rho)}{\mathrm{d}\phi}=-\frac{\mathrm{d}(r\rho)}{\mathrm{d}r}/\frac{\mathrm{d}\phi}{\mathrm{d}r}
\end{equation}

To be clear, given that we need the entire energy term to be non-negative and non-zero everywhere, we require $\frac{\mathrm{d}(r\rho)}{\mathrm{d}r}/\frac{\mathrm{d}\phi}{\mathrm{d}r}<0$. However, the term $\frac{\mathrm{d}\phi}{\mathrm{d}r}$ is always going to be positive as $\frac{\mathrm{d}\phi}{\mathrm{d}r}\equiv\frac{GM(<r)}{r^2}$ and clearly neither the radius nor the contained mass will be able to become negative. The problem comes from the other term, $\frac{\mathrm{d}(r\rho)}{\mathrm{d}r}$, i.e. the requirement that $\rho$ must fall steeper than $r^{-1}$.

\begin{figure}
\includegraphics[width=84mm]{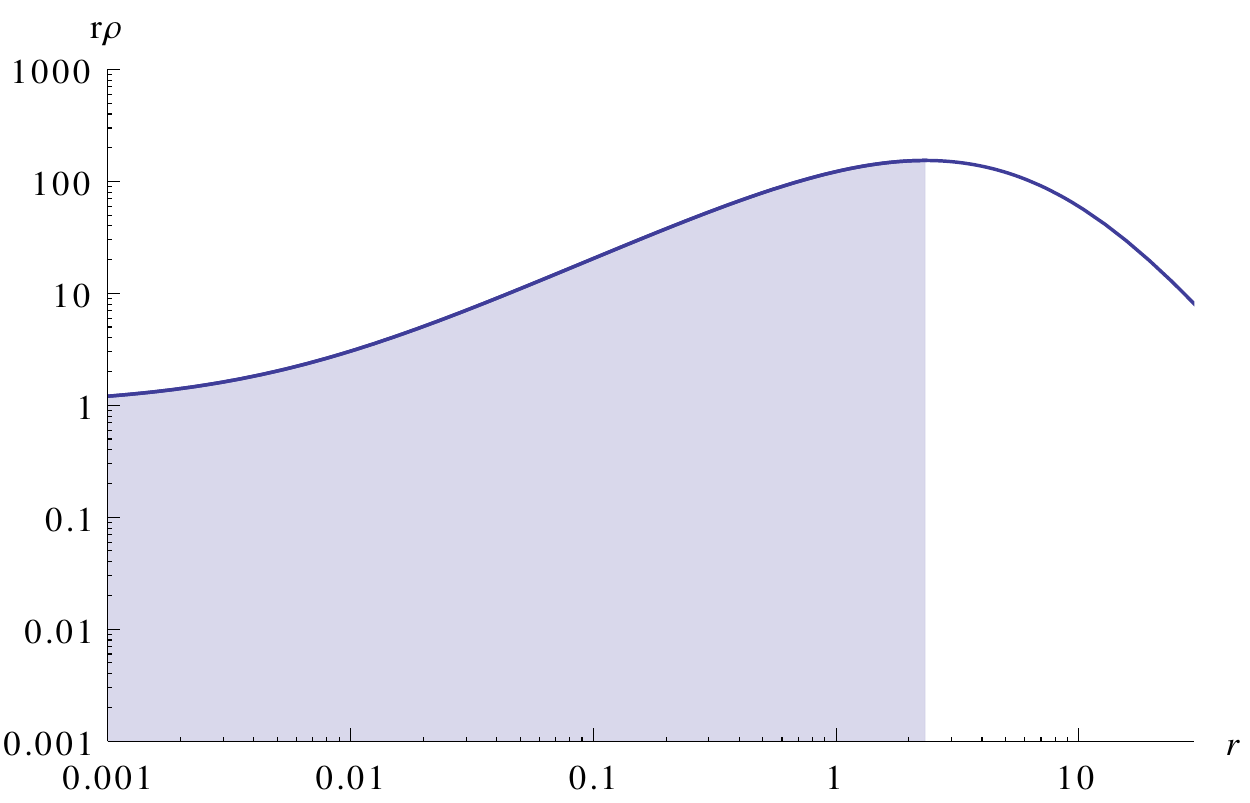}
\caption[Demonstrating the negativity of the DF]{\label{fig:jeansineqfail}Plot of $r\rho$, the differential of which comprises half of the energy function of Eq. \ref{eqn:energyfunction}. As discussed, we require this function to have a negative gradient everywhere in order for the DF to be non-negative. This figure shows that everywhere where the model could potentially fail to follow the GDSAI has a positive gradient, indicated by the shaded areas, strongly suggesting the model is unphysical.}
\end{figure}

The problem is highlighted in Fig. \ref{fig:jeansineqfail} where we see that $r\rho$ rises for all small radii, meaning the gradient is positive. This is the region that we are interested as it contains the transition between the two components of our model and the region in which we were investigating the inequality. However, since the gradient is positive here, our energy function Eq. \ref{eqn:energyfunction} will be negative. This has the unfortunate implication that there is a significant section of our system for which the DF is negative overall; \emph{the system is unphysical.}

In other words, the one region where we would might see violation of the inequality is unphysical by definition. If we are to investigate the GDSAI we are going to have to start with the DF and work up, rather than the other way around.

\section{The Distribution Function approach}
\label{sec:DF}

We set up a system where the DF is defined as:

\begin{equation}
\label{eqn:DF}
f(E,L)=A\delta(E-E_0)H\left(L^2_{cut}-L^2\right)
\end{equation}
where the constant $A$ is for dimensional consistency. This represents a system where all allowed orbits have exactly energy $\text{E}_0$ and must have angular momentum $\text{L}^2$ under $\text{L}^2_{cut}$ as defined by a delta function and a Heaviside function:

\begin{equation}
H(L^2_{cut}-L^2) = \left\{
  \begin{array}{l l}
    1 & \quad \textrm{if $L^2<L^2_{cut}$}\\
    0 & \quad \textrm{if $L^2>L^2_{cut}$}
  \end{array} \right.\
\end{equation}

This system is actually a type of polytropic model developed by \citet{Polyachenko2013} to study radial orbit instability. The model we use is equivalent to their \mbox{q=$-1$} mono-energy model and is interesting to us for its non-monotonic density profile. Given that the DF is potentially of interest, we wish to extract density and anisotropy profiles from it to examine in detail.

\subsection{Finding density}

The density is defined as the integral of the DF over all phase-space:

\begin{equation}
\label{eqn:dens}
\rho(r)=\int f(E,L)\,\mathrm{d}v_x\,\mathrm{d}v_y\,\mathrm{d}v_z
	   =\int f(E,L)\,\mathrm{d}v_r\,\mathrm{d}v_\theta\,\mathrm{d}v_\phi
\end{equation}

To ease the subsequent integration we express the integration variables in terms of E and L which we do by solving only for a constant radius, r. We use the following relationships between the velocity components:

%
%
%
%

\begin{gather}
\label{eqn:velocity}
v_\theta^2+v_\phi^2=\frac{L^2}{r^2}\\[5pt]
v_r=\sqrt{2}\sqrt{E-\Phi(r)-\frac{L^2}{2r^2}}
\end{gather}
%
%
which we then use to rewrite our integration variables:

\begin{equation}
\label{eqn:velelement2}
\mathrm{d}v_r\,\mathrm{d}v_\theta\,\mathrm{d}v_\phi=\frac{\mathrm{d}E}{v_r}\,\frac{\pi}{r^2}\,\mathrm{d}L^2
\end{equation}
%

%

We first integrate with respect to E. This is simple as there is only one function of E, namely $\text{v}_r$, and the delta function makes the integration trivial. We are then just left with the integration of $\text{L}^2$:

\begin{equation}
\rho=\frac{A\pi}{r^2}\int_0^{L^2_{cut}}  \left[2\left(E_0-\Phi-\frac{L^2}{2r^2}\right)\right]^{-\frac{1}{2}}\,\mathrm{d}L^2
\end{equation}

A similar trick is employed to deal with the Heaviside function as it has the property of constraining the limits of integration. We thus end up with an expression for the density:

\begin{equation}
\label{eqn:denfinal}
\rho=4A\sqrt{2}\pi\left( \sqrt{E_0-\Phi} - \sqrt{E_0-\Phi-\frac{L^2_{cut}}{2r^2}}\right)
\end{equation}

The second square root term can become imaginary for small values of r or particularly large values of $\text{L}^2_{cut}$. This represents parts of the system where all real orbits of the system lie under the angular momentum threshold. Consequently the only excluded orbits, which are those represented by the second term, are those which do not correspond to real possible states of the system. We avoid mathematical inconsistency in such cases by only taking the real component of the result \emph{i.e.} $0$.

\subsection{Finding anisotropy}

Next we find an expression for the anisotropy profile. We start from the definition:

\begin{equation}
\label{eqn:beta}
1-\beta(r)=\frac{\sigma_\theta^2+\sigma_\phi^2}{2\sigma_r^2}
\end{equation}

We now use the fact that $v^2\rho$ gives us the pressure along a given axis and combine it with our definition from Eq. \ref{eqn:dens}. This allows the velocity dispersion to be written as:

\begin{equation}
\label{eqn:vdispbasic}
\sigma_r^2 = \frac{\int v_r^2f(E,L)\,\mathrm{d}^3v}{\rho(r)}
\end{equation}
and likewise for $\sigma_\theta^2=\sigma_\phi^2$. We can apply this to Eq. \ref{eqn:beta} and cancel the factors of $\rho(r)$ because symmetry tells us they are equivalent. After changing variables we are left with the following:
%
%

\begin{equation}
1-\beta(r)=\frac{\int\left(\frac{L^2}{r^2}\right)f(E,L)\left(2\left[E-\Phi(r)-\frac{L^2}{2r^2}\right]\right)^{\frac{-1}{2}}\,\mathrm{d}L^2\,\mathrm{d}E}
                              {\int\left(2v_r^2\right)f(E,L)\,\mathrm{d}E\,\mathrm{d}L^2}
\end{equation}

As before we easily perform the integration over E and use the Heaviside functions to place limits on the integration over $\text{L}^2$ but the resulting expression is more complicated:

\begin{equation}
1-\beta(r)=\frac{\frac{1}{\sqrt{2}}\int^{L^2_{cut}}_0\left(\frac{L^2}{r^2}\right)\left(E_0-\Phi(r)-\frac{L^2}{2r^2}\right)^{\frac{-1}{2}}\,\mathrm{d}L^2}
                              {2\sqrt{2}\int^{L^2_{cut}}_0\left(E_0-\Phi(r)-\frac{L^2}{2r^2}\right)^{\frac{1}{2}}\,\mathrm{d}L^2}=\frac{1}{4}\frac{\mathbb{I}_1}{\mathbb{I}_2}
\end{equation}

These integrals can be solved analytically by the substitutions shown in the appendix and the result for $\beta$ expressed as:

\begin{equation}
\label{eqn:ani}
\beta=\frac{\sqrt{1-x}\left(\frac{3x}{2}\right)}{1-(1-x)^{\frac{3}{2}}}\text{ where }x=\frac{L^2_{cut}}{2r^2\left(E_0-\Phi\right)}=\frac{L^2_{cut}}{r^2v^2_c}
\end{equation}

As with the density profile the anisotropy profile can produce imaginary results if $x>1$ in regions where \mbox{$\frac{L^2_{cut}}{r^2}>\frac{L^2_{max}}{r^2}$}. This, again, corresponds to regions of the system where all real orbits lie below the angular momentum threshold. Since we know that this case is not physical we resolve it by enforcing a maximum value such that $\forall x>1\text{, }x=1$. This constrains the angular momentum threshold to be locally no greater than the largest possible angular momentum at radii where the imaginary numbers would otherwise be produced. This does not change the physical implications of the formula and just ensures mathematical consistency.

We can calculate $\sigma_r^2$ by using Eq. \ref{eqn:vdispbasic} and following a similar line of reasoning as for the anisotropy. Using the parameterisation from Eq. \ref{eqn:ani} yields:

\begin{equation}
\sigma_r^2=\frac{2\left(E_0-\Phi\right)\left[1-\left(1-x\right)^{\frac{3}{2}}\right]}{9\left(1-\sqrt{1-x}\right)}
\end{equation}

The tangential dispersions can either be found similarly or by combining the radial dispersion with the anisotropy. They won't be reproduced here as they are not used in our analysis.

\subsection{Finding potential}

We can use Poisson's equation to find the potential by integrating the following second order ODE:

\begin{equation}
\label{eqn:poissonfinal}
\frac{\mathrm{d}^2 (r  \Phi) }{ 4 \pi G r\,\mathrm{d}r^2}=4\sqrt{2}\pi A \left( \sqrt{E_0-\Phi} - \sqrt{E_0-\Phi-\frac{L^2_{cut}}{2r^2}}\right)
\end{equation}
where, to be rigorous, we have included the dimensional pre-factor $A$ from our DF of Eq. \ref{eqn:DF}. This can be cast into a dimensionless form by the following scaling:

\begin{equation}
\label{eqn:poissonfinal2}
 r= B \tilde{r}\text{; }\Phi =   E_0 -  C \tilde{\Psi} \rightarrow \frac{\mathrm{d}\left(E_0-\tilde{\Psi}\right)}{\mathrm{d}\tilde{r}}\equiv-\frac{\mathrm{d}\tilde{\Psi}}{\mathrm{d}\tilde{r}}\text{; }L^2=B^2C\tilde{L}^2
\end{equation}
where we require that $C=(AB^2G)^2$ for consistency. This gives us the following dimensionless expression that requires solving:

\begin{equation}
 -\frac{1}{4\pi\tilde{r}^2}\frac{\mathrm{d}}{\mathrm{d}\tilde{r}}\left(\tilde{r}^2\frac{\mathrm{d}\tilde{\Psi}}{\mathrm{d}\tilde{r}}\right) = 4\sqrt{2}\pi \left(\sqrt{\tilde{\Psi}} - \sqrt{\tilde{\Psi}-\frac{\tilde{L}^2_{cut}}{2\tilde{r}^2}}\right)=4\sqrt{2}\pi\tilde{\rho}(\tilde{r})
\end{equation}

As before we have an non-physical case for small r and large $\text{L}^2_{cut}$ which we resolve by taking only the real part of the root. To solve the ODE we apply standard initial conditions that:

\begin{equation}
\label{eqn:conditions}
\tilde{\Psi}(\tilde{r}=0)=(4\pi)^2\text{; }\frac{\mathrm{d}\tilde{\Psi}(\tilde{r}=0)}{\mathrm{d}\tilde{r}}=0
\end{equation}
where the choice of the constant $(4\pi)^2$ is arbitrary and has been chosen here to make the radius of the system of order unity. Regrettably this equation must be solved numerically and the resulting $\tilde{\Psi}(r)$ is shown in Fig. \ref{fig:ineqpotplot}.

\section{Understanding the system}
\label{sec:result}

We have arrived at an analytically self-consistent system which should avoid the problems from \S\ref{sec:jeans}. We can summarise the system by collating our results so far:

\begin{gather}
\label{eqn:system}
\rho=4\sqrt{2}\pi A\left(\sqrt{\Psi}-\sqrt{\Psi-\frac{L^2_{cut}}{2r^2}}\right)\nonumber \\[5pt]
\sigma_r^2=\frac{2\left(E_0-\Phi\right)\left[1-\left(1-x\right)^{\frac{3}{2}}\right]}{9\left(1-\sqrt{1-x}\right)}\nonumber \\[5pt]
\beta=\frac{\sqrt{1-x}\left(\frac{3x}{2}\right)}{1-(1-x)^{\frac{3}{2}}}\text{ where }x=\frac{L^2_{cut}}{2r^2\left(E_0-\Phi\right)}\nonumber
\end{gather}

\subsection{Characterising the density profile}

We first of all note that, for a particular value of $\text{L}_{cut}$  our density profile is an augmented density profile \emph{i.e.} our density can be expressed only in terms of $\Psi$ and $r$. However, unlike the augmented density functions examined in \citet{Ciotti2010} and \citet{vanhese2011}, ours is not separable in terms of those variables. This means that our model falls outside the set of models for which the GDSAI has been studied. However, our model is also highly unusual and possess profiles that are distinctly artificial so we will spend this section characterising and explaining the model before drawing any conclusions.

The density at a given point can be thought of as the amount of orbits which require a particle to pass through that radius. As we established, our DF from Eq. \ref{eqn:DF} means that we are only allowing orbits with angular momentum $0<\text{L}^2<\text{L}^2_{cut}$ and energy $\text{E}=\text{E}_0$. It is convenient to think of the density as being \emph{the sum of all possible, physical orbits of energy $\text{E}_0$ minus all orbits of energy $\text{E}_0$ and $\text{L}^2>\text{L}^2_{cut}$ that pass through a certain radius}. We can see this interpretation directly in Eq. \ref{eqn:denfinal} where the density is the difference between two terms which, as we will now discuss, correspond to the description above.

\begin{figure}
\includegraphics[width=84mm]{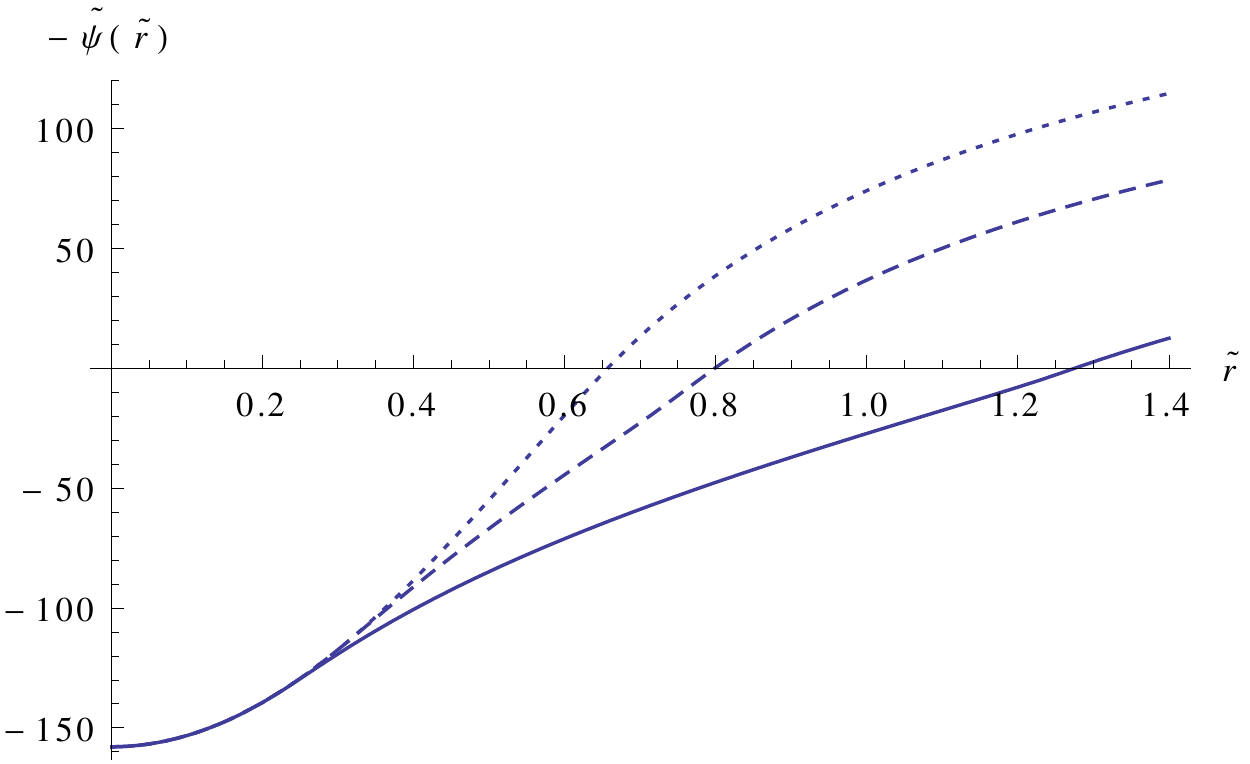}
\caption[Numerical potential from our DF]{\label{fig:ineqpotplot}The numerically derived potential for our distribution function. The potential becomes positive at the point where $\tilde{\rho}=0$. The potential is plotted for three models corresponding to \mbox{$\xi=\{1,1.5,2\}$} (see Eq. \ref{eqn:xi}) for the solid, dashed, and dotted lines respectively. Although $\tilde{\Psi}$ is decreasing faster than $\tilde{r}$ in the outer regions, it will behave like a Kepler potential at radii larger than the size of the system as the system is truncated at finite radius.}
\end{figure}

The first term, $\sqrt{E_0-\Phi}$, represents all physical orbits of energy $\text{E}_0$. As we can see from Fig. \ref{fig:ineqpotplot}, the potential of the system is a monotonically increasing function.

The second term, $\sqrt{E_0-\Phi-\frac{L^2_{cut}}{2r^2}}$, is more complex due to the addition of an angular momentum term. This represents all orbits of energy $\text{E}_0$ and an angular momentum of at least $\text{L}^2_{cut}$.

In Fig. \ref{fig:ineqpotplot}, and in most subsequent figures, we plot a handful of models with different values of $\tilde{\text{L}}^2_{cut}$ as this is key parameter which determines the behaviour of the model. The models are generated by:

\begin{equation}
\label{eqn:xi}
\tilde{L}^2_{cut}=14.37\xi
\end{equation}
where $\xi$ is a free parameter and useful index for a particular model. The constant $14.37$ was chosen to give an arbitrary but convenient value for $\tilde{\text{L}}^2_{cut}(\xi=1)$ and is related to $\tilde{\Psi}(\tilde{r}=0)$. As we can see from Fig. \ref{fig:ineqpotplot}, increasing $\tilde{\text{L}}^2_{cut}$ will decrease the radius of the system. To allow for better comparison of models we normalise the radius of each model to 1  in subsequent figures.

\begin{figure}
\includegraphics[width=84mm]{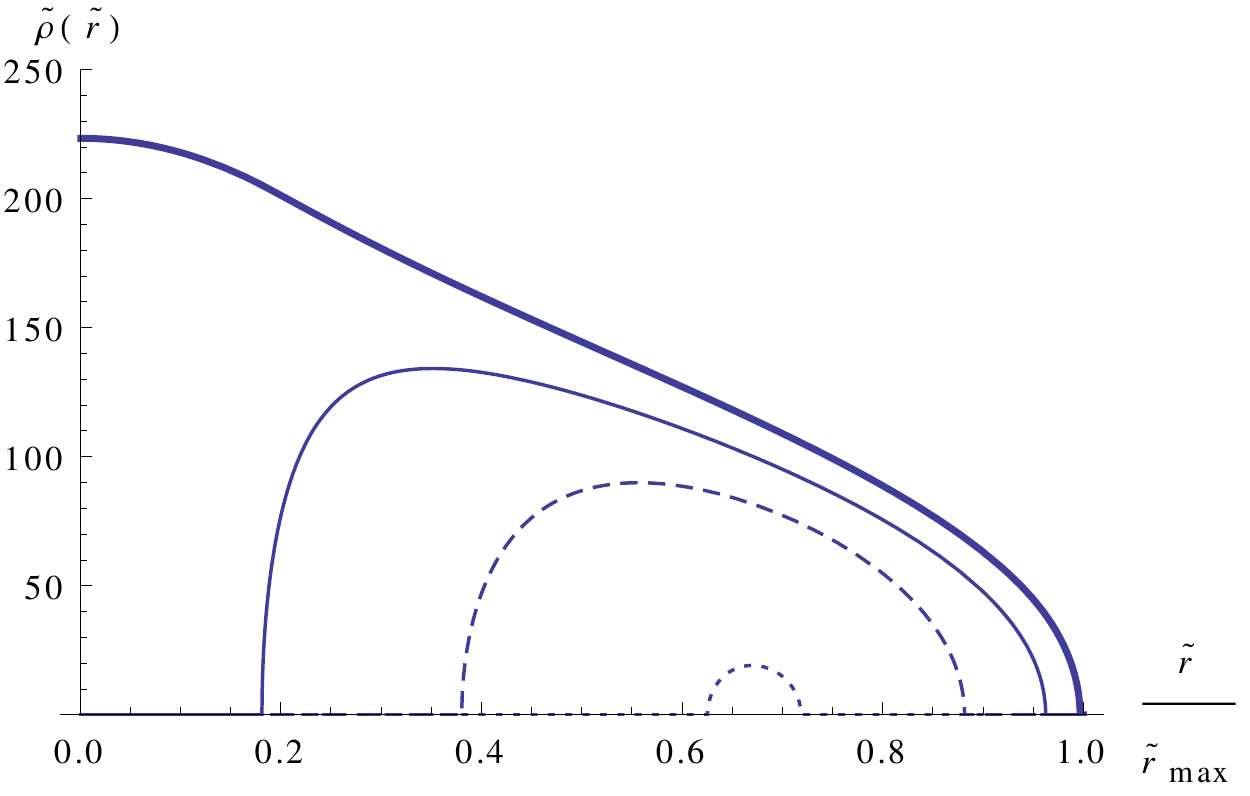}
\caption[The components of the DF density profile]{\label{fig:ineqsepdenplot}Both components of the density in our example model. The area under the largest curve (thick, solid line) contains all physical orbits in the system while the smaller curves (solid, dashed and dotted) contain all allowed orbits that have $\text{L}^2$ \emph{greater} than $\text{L}^2_{cut}$ for \mbox{$\xi=\{1,1.5,2\}$} (see Eq. \ref{eqn:xi}).}
\end{figure}

As we see from Fig. \ref{fig:ineqsepdenplot} each term in the density can itself describe a meaningful denisty and we have arranged them such that the number of allowed orbits is proportional to the area under the curve. The area under the largest curve contains all orbits given by $\sqrt{E_0-\Phi}$ and any physically permitted angular momentum. The area under the smaller curve is given by $\sqrt{E_0-\Phi-\frac{L^2_{cut}}{2r^2}}$ which only contains orbits with angular momentum \emph{greater} than $\text{L}^2_{cut}$ for several different values of the threshold.

\begin{figure}
\includegraphics[width=84mm]{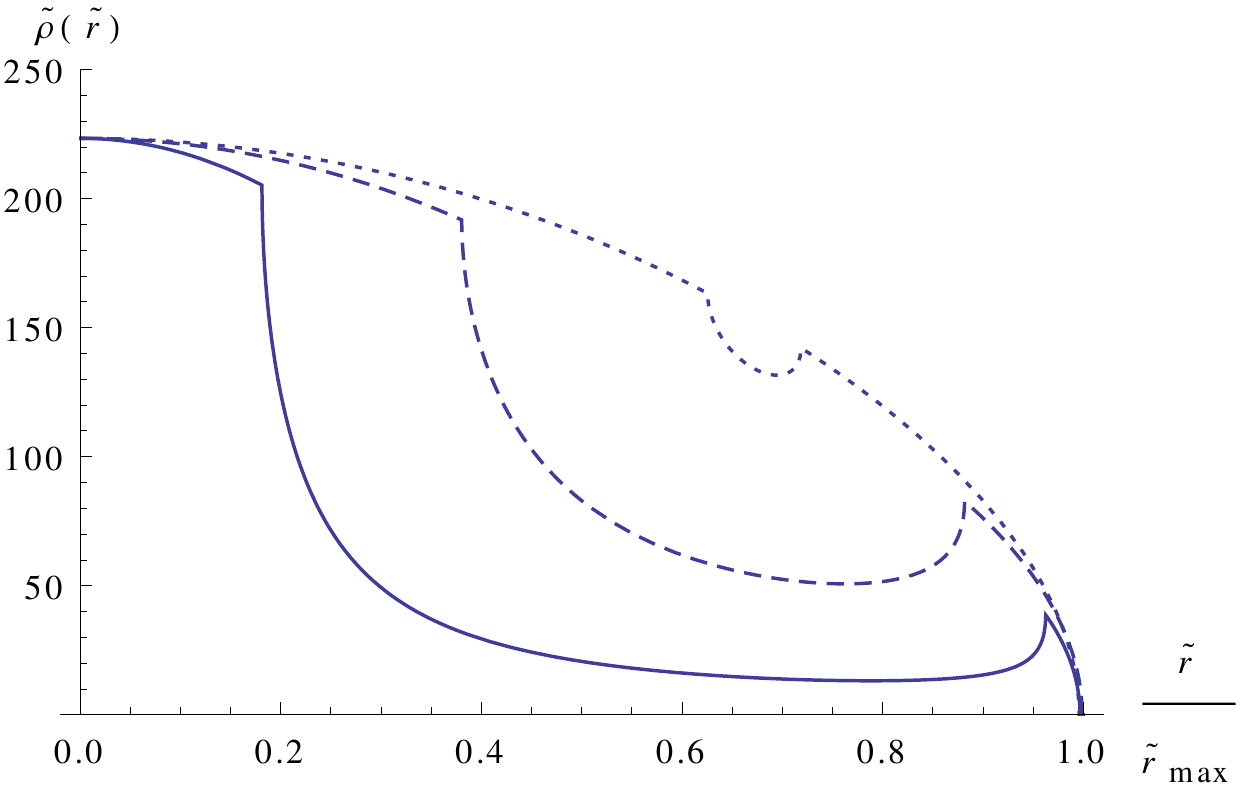}
\caption[The density profile of our DF]{\label{fig:ineqhorndenplot}The full density profile for our distribution function. The density is plotted for three models corresponding to \mbox{$\xi=\{1,1.5,2\}$} (see Eq. \ref{eqn:xi}) for the solid, dashed, and dotted lines respectively. Significant violation of the GDSAI is expected to occur in the sharp peak at large radii.}
\end{figure}

We recall that our overall density is the difference between these two components which produces Fig. \ref{fig:ineqhorndenplot}. The immediate feature of note is the sharp peak towards the outer edge which is the feature that makes this system so useful.

Now that we understand how this profile is formed we explain why these features arise. It must be borne in mind that this is a mono-energy system and that all orbits have total energy of exactly $\text{E}_0$.

At small radii, our angular momentum constraint has minimal impact. The small value of r means that even a highly tangential orbit will have an angular momentum that is under the $\text{L}^2_{cut}$ threshold. We thus find there are a large number of possible orbits and the density is high.

As we move further out an orbit of given circularity will have higher angular momentum so allowed orbits will be progressively more radially anisotropic in order to fit under the angular momentum limit. Accordingly, fewer allowed orbits exist at these radii and the denisty is lower than expected.

\begin{figure}
\includegraphics[width=84mm]{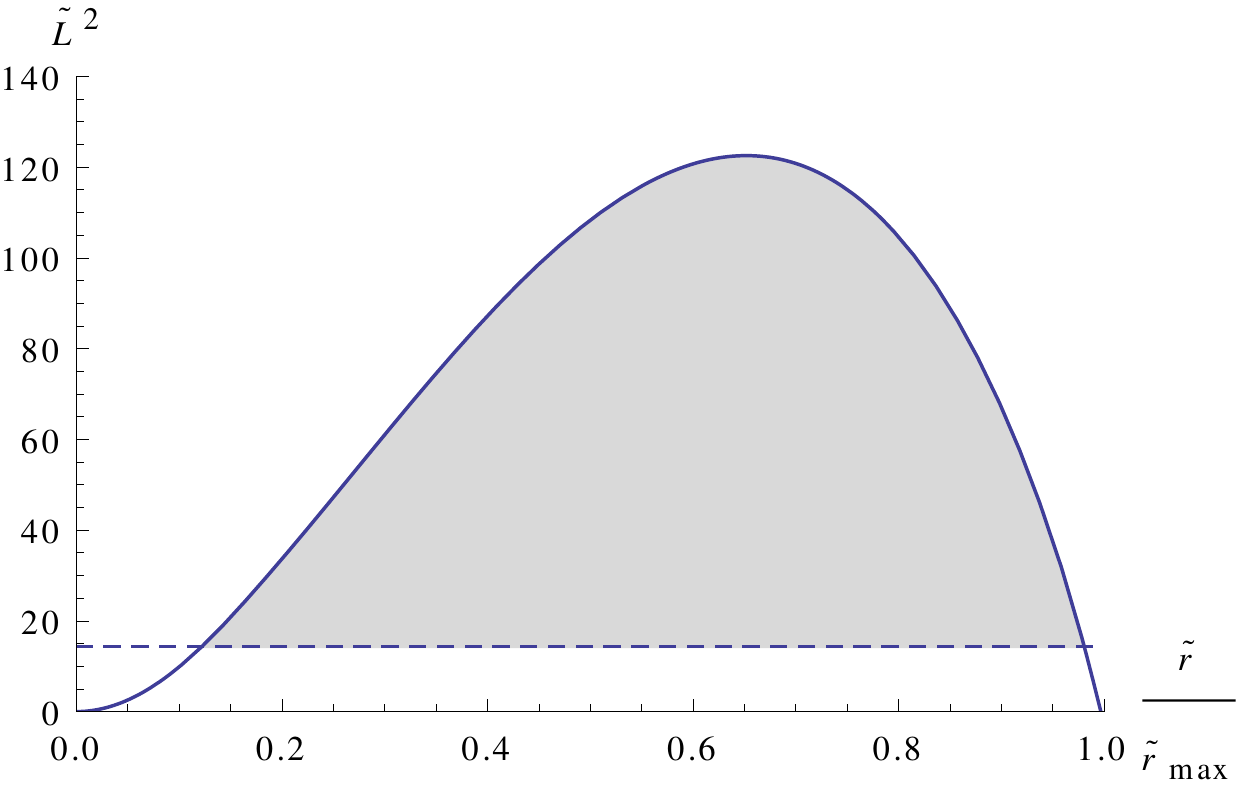}
\caption[Angular momentum of circular orbits in our DF]{\label{fig:ineqlplot}Variation of the maximum possible angular momentum, $\tilde{\text{L}}^2_{max}(\tilde{r})$, with radius showing that at large and small radii it is impossible for a particle of energy $E_0$ to surpass the angular momentum cutoff. The cutoff $\tilde{\text{L}}^2_{cut}(\xi=1)$ is indicated by the horizontal line with the shaded region representing otherwise permissible orbits that will fail the angular momentum cut. $\tilde{\text{L}}^2_{max}$ was found by using Eq. \ref{eqn:lmax}.}
\end{figure}

Finally, in the outermost regions the angular momentum limit no longer has any impact. Although the radius is large, the angular momentum of a highly tangetnial orbit here is very low as the majority of a particle's energy is used to overcome the potential. This means that little is left for kinetic energy and thus the tangential velocity is extremely small. Even a completely circular orbit at this radius will have an angular momentum below the threshold. Using the same logic as for small radii this means that the density will be proportionally higher as all the possible orbits are permitted by the threshold. This can be seen in Fig. \ref{fig:ineqlplot}.

The last feature to understand is the sharp rise in the density at large radii. This is difficult to discuss analytically due to the lack of an analytical solution to the potential. For example, if we try to understand how angular momentum changes with radius we could look at the angular momentum of an orbit whose apocentre is at a given radius. If the particle is at its apocentre then we know that $v_r=0$ at that point and that the angular momentum will be:

\begin{equation}
\label{eqn:lmax}
L^2_{max}=2r^2\Psi(r)
\end{equation}

Since we do not know the dependence of $\Psi$ on r we cannot construct analytical expressions for the slope. Thus, to explain the reason for the increase in density we must rely on the figures and numerical results to support the explanation.

At the point where the density rises we can see from Fig. \ref{fig:ineqlplot} that $\text{L}^2_{max}$ is decreasing rapidly. This means that, as the radius increases, a particle can eventually have an increasingly large tangential velocity at apocentre and still be under the angular momentum threshold. In other words, a particle is allowed to make a larger angle between the radial axis and its velocity vector the further it is from the centre. This also means that the amount of orbits possible at these radii, \emph{i.e.} the density, is increasing in proportion.

In the absence of an angular momentum limit then, as seen in Fig. \ref{fig:ineqsepdenplot}, the density naturally decreases monotonically as the negative potential increases towards the edge of the system. However, comparing these two figures shows that the rate of \emph{decline} in phase-space density due to the potential is \emph{lower} than the rate of \emph{increase} due to the range of allowed angles for velocity vectors. In other words, the increase in density due to the lessening impact of $\text{L}_{cut}$ overpowers the natural decline in phase-space density due to the potential.

This means that over a small range of radii the density actually increases until all orbits fall under the angular momentum threshold again. At this point the system has no orbits left to be added to the density as the radius increases because none are being excluded and the decline in density resumes until the edge of the system.

\subsection{Characterising the anisotropy profile}

\begin{figure}
\includegraphics[width=84mm]{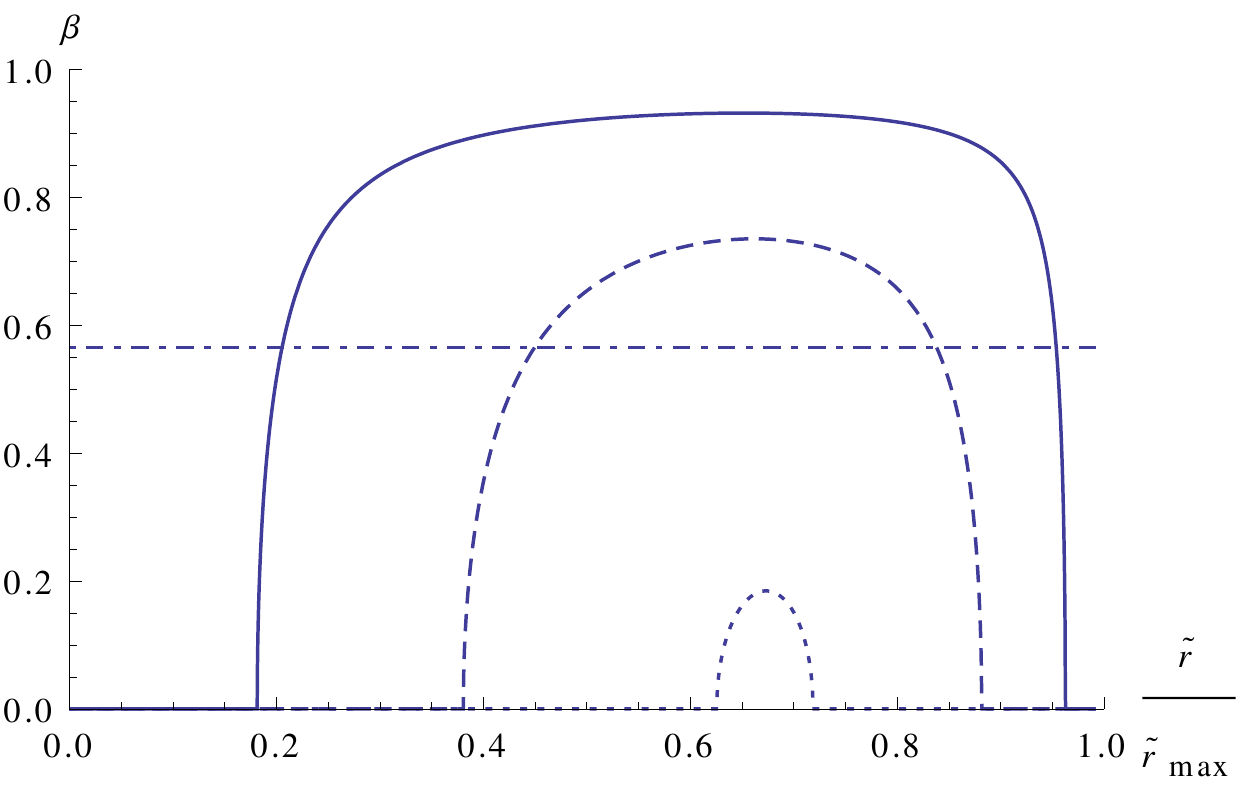}
\caption[Anisotropy profile of our DF]{\label{fig:ineqbetaplot}The anisotropy profile for our system demonstrating the isotropic core and edge regions for the three models $\xi=\{1,1.5,2\}$ (the solid, dashed, and dotted curves). The dot-dashed line represents the anisotropy corresponding to the nominal radial orbit instability threshold of $2\bar{\text{T}}_r/\bar{\text{T}}_t\approx2.3$ \citep{MerrittAguilar1985}.}
\end{figure}

One key feature of this model is that it has very low central anisotropy. The examination of \citet{vanhese2011} proved the GDSAI held for all separable systems with \mbox{$\beta_0\leq1/2$} which makes an investigation into non-separable models with $\beta_0=0$ of particular interest. In fact our models are isotropic at both small and large radii and only become radially anisotropic for a set of intermediate radii as shown in Fig. \ref{fig:ineqbetaplot}. 

As discussed in our consideration of the density profile, the angular momentum threshold removes no orbits from the core of the system or the outskirts. This is because even particles at apocentre at these radii have low angular momentum due to either the small radius of the orbit or the low tangential velocity of the particle. At these radii the angular momentum function of our DF is fixed at $H\left(L^2_{cut}-L^2\right)=1$ for all orbits and thus the total DF is a function only of energy. This means that all energy is necessarily split evenly between velocity components and that region is isotropic \citep{BinneyTremaine}.

In the regions where the angular momentum limit \emph{is} removing orbits, the change in anisotropy can be thought of as follows. Imagine trying to construct a particle on an orbit that tries to maximise its angular momentum by minimising its radial velocity component, as we did when constructing Eq. \ref{eqn:lmax}. Since this orbit must have a certain amount of energy it has a very predictable angular momentum which will put the particle over the $\text{L}^2_{cut}$ threshold. This means that this orbit, and all highly tangential ones, are removed, leaving only the more radial ones. Given that the system would be isotropic if not for this process, we can say that any radii at which the angular momentum cut removes orbits is guaranteed to be radially anisotropic.

This behaviour is shown directly by looking again at Fig. \ref{fig:ineqlplot}. The shaded orbits above the angular momentum threshold $\tilde{\text{L}}^2_{cut}$ are excluded which implies that any radii at which a portion of the area under the curve is shaded will be radially anisotropic. The amount of anisotropy will grow with the size of the shaded area.

\begin{figure}
\includegraphics[width=84mm]{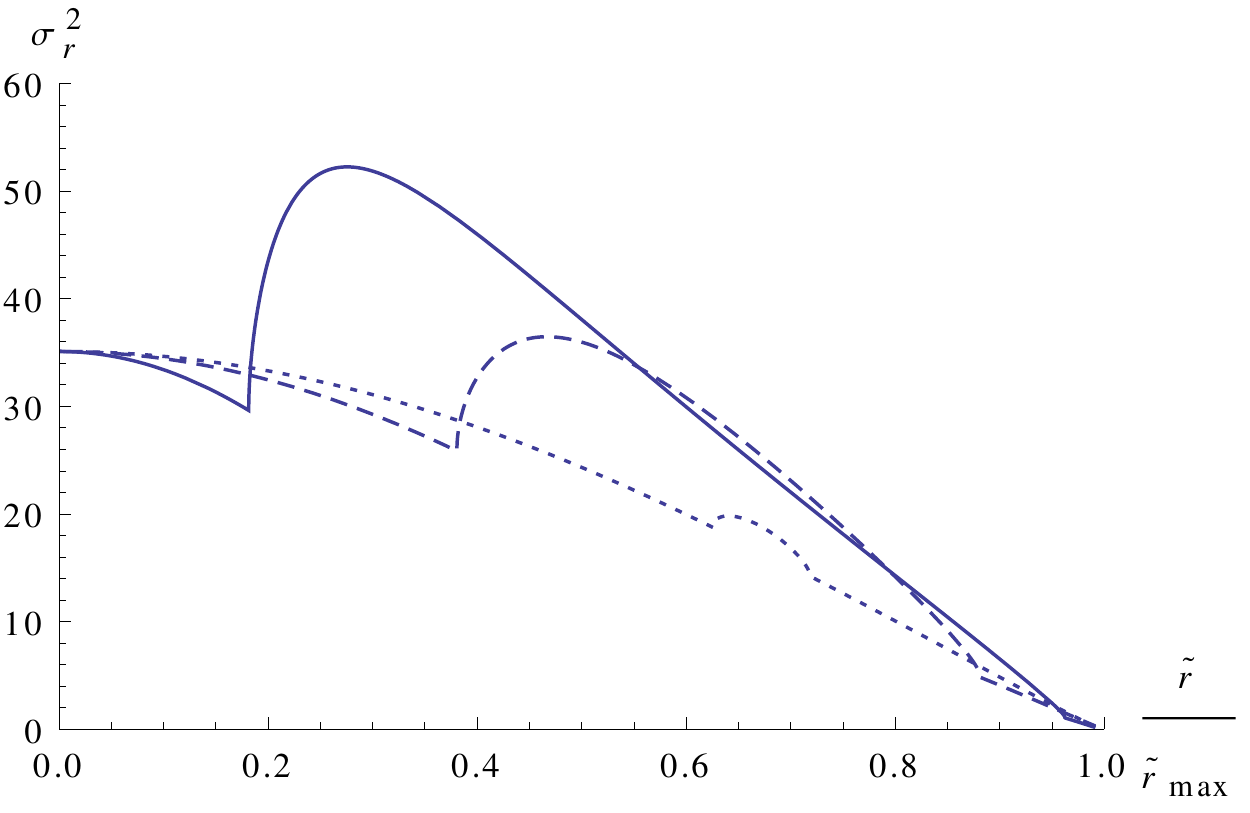}
\caption[Radial velocity profile of our DF]{\label{fig:ineqsigmaplot} The radial velocity dispersion profile of our system for the three models \mbox{$\xi=\{1,1.5,2\}$} (the solid, dashed, and dotted curves). Upon entering or leaving regions of the model where the angular momentum cut is removing orbits there is a discontinuity as there is a abrupt decrease in the amount of energy that can be used in tangential motion.}
\end{figure}

Because the transition between the isotropic regions and the anisotropic regions is a sharp one we find discontinuities in the gradient of individual velocity dispersions. Fig. \ref{fig:ineqsigmaplot} shows how the slow decrease of energy in the radial velocity component is sharply reversed upon reaching the anisotropic regions of the system. A smaller discontinuity is also present upon reaching the isotropic regions at large radii. These correspond exactly to the two discontinuities in the slope of the density profile.

\subsection{The inability to extend the GDSAI to this model}

We can now confirm that our system does not obey a GDSAI-like relation at certain radii. We have proved that our system is radially anisotropic in the regions where the angular momentum threshold is removing orbits. We have also demonstrated that our system's density profile is either flat or rising in those same regions due to the exclusion of a region of phase-space. From this we can see that there are two regions where we would expect to find that $\gamma<2\beta$ which runs counter to an attempt to extend the GDSAI.

At the radii immediately prior to the density peak we are guaranteed not to recover a relationship that follows the GDSAI as the density is \emph{rising} sharply with radius and so $\gamma<0$. However, we also fail to find the relation at a large range of intermediate radii where the density profile is approximately flat and thus $\gamma\approx 0$. Since the anisotropy here is high, the system can be configured such that $\gamma<2\beta$ at these radii as well.

\begin{figure}
\includegraphics[width=84mm]{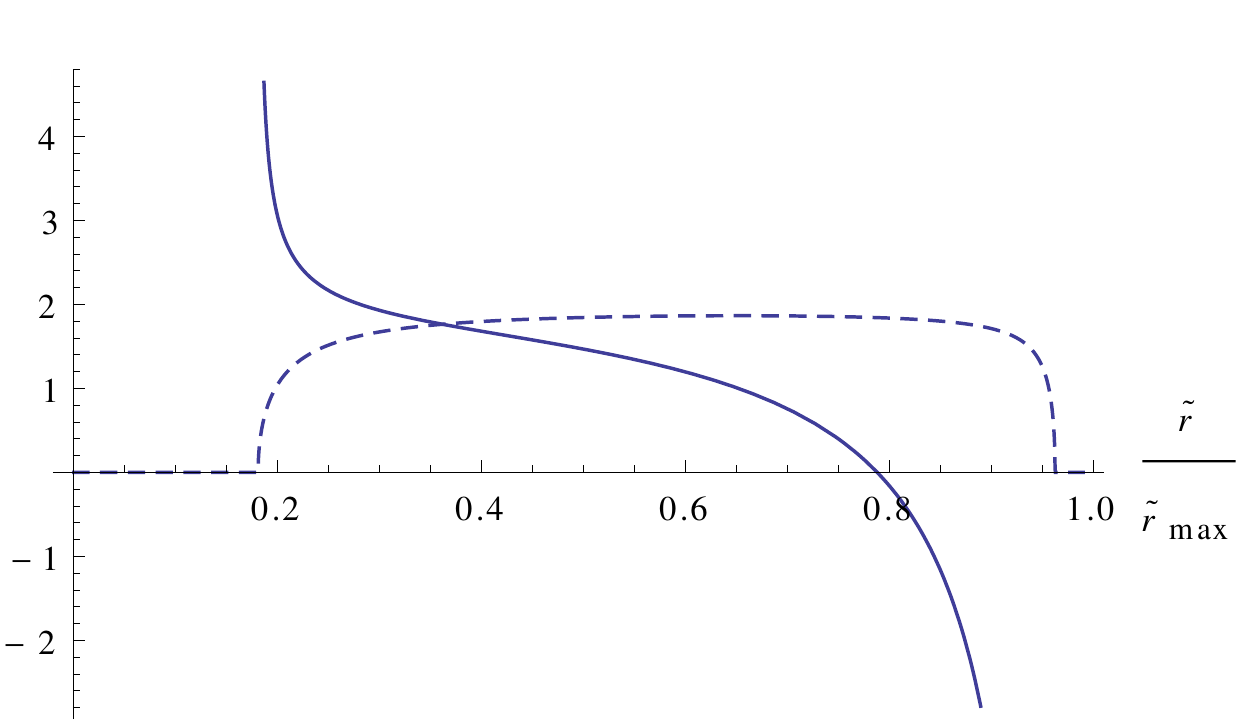}
\caption[Demonstrating GDSAI violation in our DF]{\label{fig:violationplot}Showing both $\gamma$ and $2\beta$ as functions of radius for the model $\xi=1$. The point where the lines cross represents the beginning of the regions that fail to obey a GDSAI-like relation.}
\end{figure}

The degree to which our system fails to follow the same inequality as the GDSAI and the regions in which this occurs are plotted in Fig. \ref{fig:violationplot} where we show curves of $\gamma$ and $2\beta$ for the model $\xi=1$. Any radius where $\gamma<2\beta$ demonstrates that the GDSAI could not be extended to include this model. We can see that approximately 2/3 of radii in this model display such behaviour, corresponding to $16\%$ of the model's mass. Accordingly, we suggest that the GDSAI cannot be extended to guarantee the existence or non-existence of phase-space consistency in a non-separable DF of this kind.

Having demonstrated the theoretical interest of our system we will now discuss the stability of the equilibrium solution found for our DF.

\section{System stability}
\label{sec:data}
\subsection{Radial Instability}

A fundamental measure of radial stability is the anisotropic extension of the Doremus-Feix-Baumann theorem \citep{DoremusFeixBaumann1971,DoremusFeixBaumann1973, Doremusetal1976,BinneyTremaine} which states that a stable system must satisfy d$f_0$/dH$_0<0$. For our DF this requires that d$f$(E)/dE$<0$ which is problematic because $f(\text{E})=\delta(\text{E}_0-\text{E})$ and, due to the peculiarities of the Dirac delta, its derivative is formally undefined. However, since this is also a necessary criteria for the emergence of the H\'{e}non Instability \citep{BarnesGoodmanHut1986,Merritt1999} we can use the H\'{e}non criteria as an indicator of potential radial instability.

\subsection{H\'{e}non Instability}
\label{sec:Henon}

Perhaps the most similar system to ours to undergo extensive stability testing is the $n=1/2$ polytrope. The testing of \citet{Henon1973} and \citet{BarnesGoodmanHut1986} demonstrated that the oscillatory stability of the polytrope was due a uneven radial velocity distribution that was termed the `H\'{e}non Instability'. It is interesting to note that the systems of \citet{vanhese2011} which demonstrated that the GDSAI lacked predicitive power for separable systems of $\beta_0>1/2$ were unstable according to the H\'{e}non criteria.

The H\'{e}non instability will appear in our systems if they possess two or more distinct peaks in the radial velocity distribution $P(v_r)$ where:

\begin{gather}
P(v_r)=\iint^{+\infty}_{-\infty}\delta\left(E-E_0\right)H\left(L^2_{cut}-L^2\right)\mathrm{d}v_\theta \mathrm{d}v_\phi\nonumber\\[5pt]
\quad\quad\quad\quad\quad{}=H\left(\frac{1}{2}v^2_r-\Psi(r)+\frac{L^2_{cut}}{2r^2}\right)-H\left(\frac{1}{2}v^2_r-\Psi(r)\right)
\end{gather}

This distribution will, after being normalised, give us a probability density that is constant over a narrow range of radial velocities:

\begin{equation}
P(v_r)\neq0\text{ where }\frac{-L^2_{cut}}{2r^2}<\frac{1}{2}v^2_r-\Psi(r)<0
\end{equation}

\begin{figure}
\includegraphics[width=84mm]{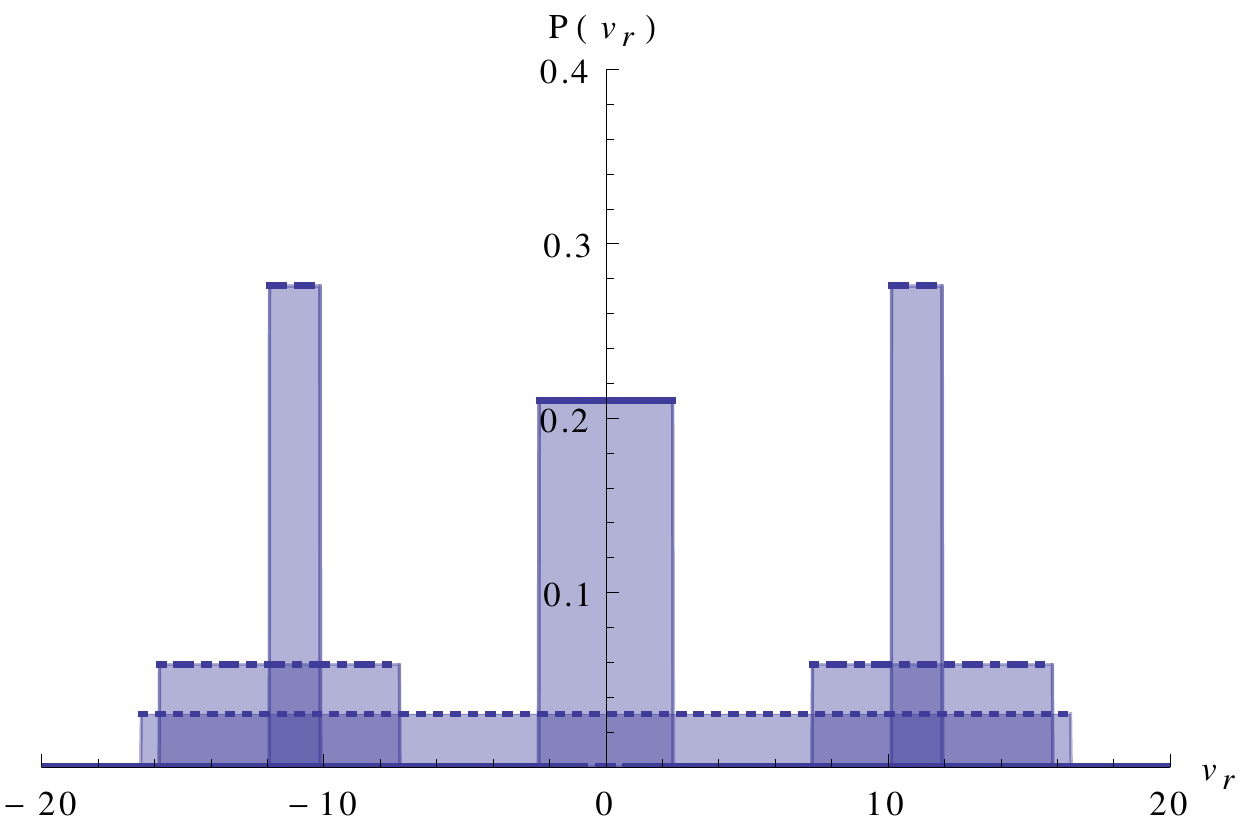}
\caption[H\'{e}non instability in our DF]{\label{fig:henonplot}A probability distribution bar plot showing the normalised probability of finding a particle with a given radial velocity at certain radii. Bars topped with a dotted, dot-dashed, dashed, and solid line represent the probabilities at $\tilde{r}=\{0.22,0.27,0.6,1.25\}$ respectively.}
\end{figure}

Between this and our understanding about the allowed orbits we can explain the distribution seen for a selection of radii for the model $\xi=1$ in Fig. \ref{fig:henonplot}.

At small radii, $\frac{-\tilde{L}^2_{cut}}{2\tilde{r}^2}\ll0$ so the only constraint on $v_r$ is energy conservation. Thus we expect a range of velocities out to some maximum value.

At intermediate radii a particle is not allowed to have negligible radial velocity. If a particle here has $v_r\approx0$ then the mono-energy constraint would demand that it compensate with significant $v_t$ which, at these radii, would put it over the $\tilde{\text{L}}^2_{cut}$ threshold. Accordingly, the radial velocity distribution at these radii will be two sharp peaks with a gap around $v_r=0$. The width of the peaks is determined by $\tilde{\text{L}}^2_{cut}$ and the radius.

Finally, at the outer edge of the system the angular momentum threshold no longer removes orbits and so particles with $v_r\approx0$ are allowed once again. The two peaks reform into a single peak centred around $v_r=0$ like at small radii but with a smaller width due to the smaller amount of kinetic energy available at these radii.

Since our models clearly possess two very sharp and well-defined peaks at most radii we conclude that this model may be susceptible to the H\'{e}non instability.

\subsection{Radial Orbit Instability}

\begin{figure}
\includegraphics[width=84mm]{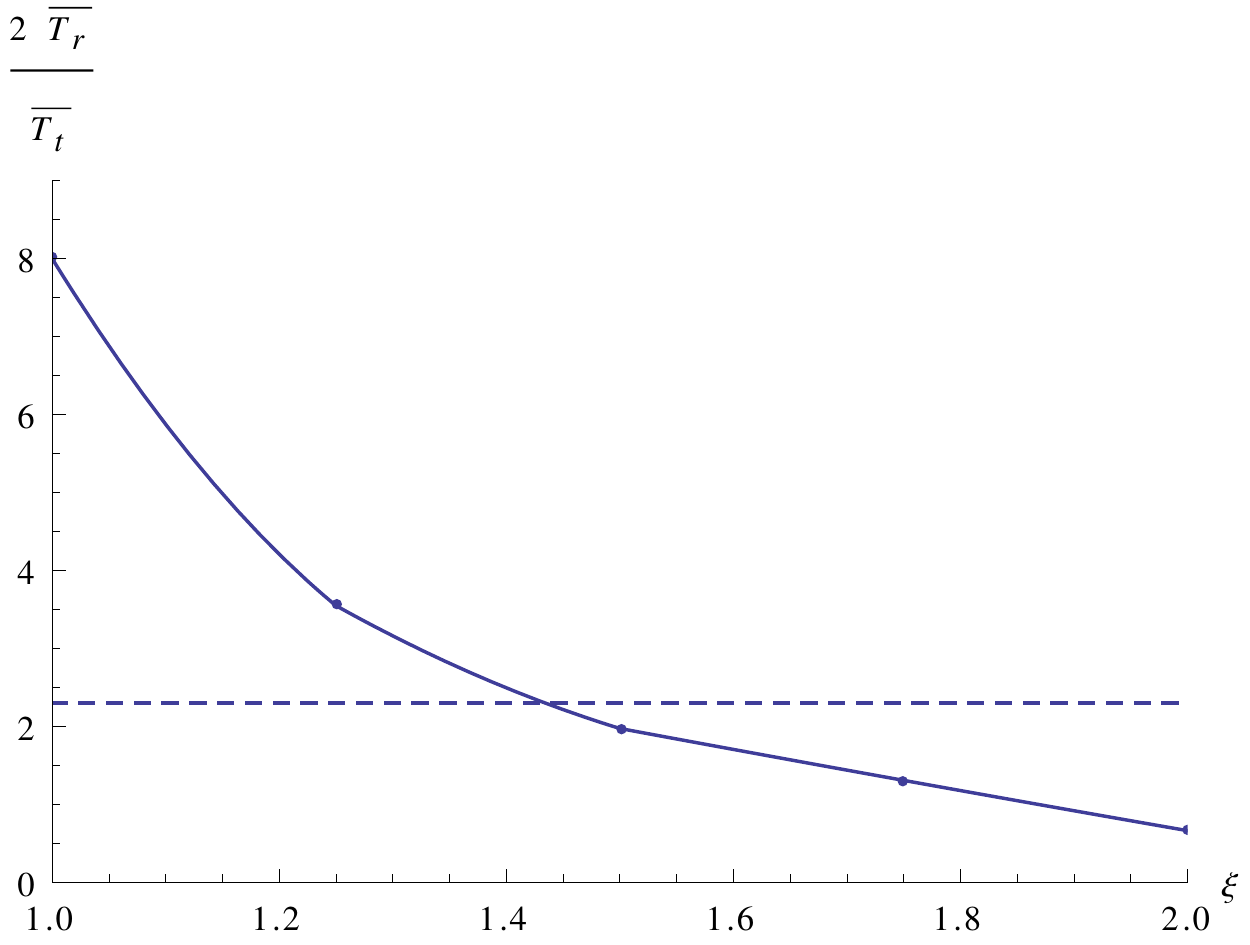}
\caption[ROI as a function of angular momentum threshold]{\label{fig:roicriteriaplot}The dependence of  $2\bar{\text{T}}_r/\bar{\text{T}}_t$ against the $\xi$ of the model. The radial orbit instability criteria of  $2\bar{\text{T}}_r/\bar{\text{T}}_t\approx2.3$ is marked with a dashed line for clarity. The stability of the model depends on $\xi$ with larger values producing more stable systems.}
\end{figure}

The stability of our system to non-radial modes is easier to assess. We use the simple stability measure of $2\bar{\text{T}}_r/\bar{\text{T}}_t=2.3$ which has been the subject of much debate \citep{Merritt1999}. Looking back at Fig. \ref{fig:ineqbetaplot} we see that the model $\xi=1$ is over the limit throughout most of the system. However, models with a higher angular momentum threshold can consistently have low enough anisotropy to avoid the instability.

This is shown in Fig. \ref{fig:roicriteriaplot} where the radial orbit instability criteria is plotted against the angular momentum threshold. Models with $\xi\approx1.45$ and above appear to be stable. So, it appears that the stability of the system is a function of the parameter $\xi$.

\subsection{Stability dependence on $\text{L}^2_{cut}$}

We have seen how our system has the potential to suffer from a variety of stability problems. However, there is reason to believe that the degree of instability can be controlled if not mitigated entirely.

We begin by once again noting that in regions where $\text{L}^2_{cut}$ is higher than the largest possible angular momentum (see Fig. \ref{fig:ineqlplot}) our system behaves as if the DF is exclusively a function of energy and is thus always isotropic. Additionally, we consider that $\text{L}^2_{cut}$ is a tunable parameter through Eq. \ref{eqn:xi}.

\begin{figure}
\includegraphics[width=84mm]{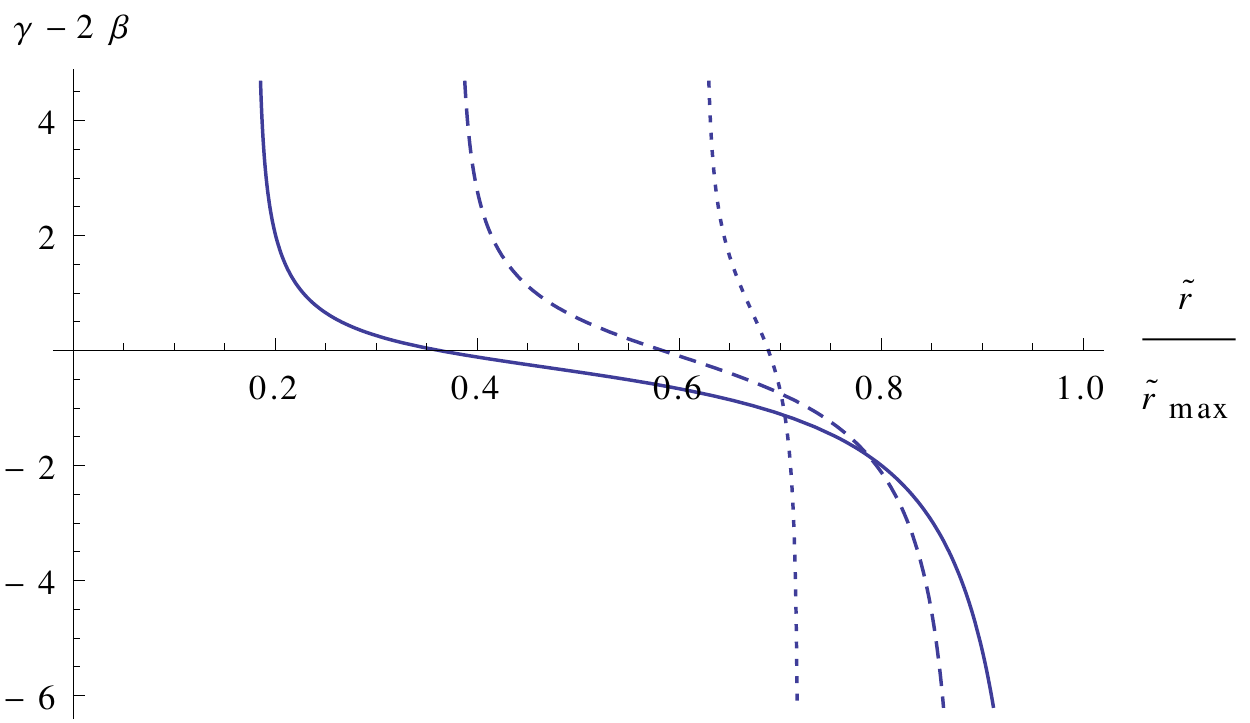}
\caption[GDSAI function of our system]{\label{fig:hornchangel2plot}The GDSAI function plotted as $\gamma-2\beta$ for the three models $\xi=\{1,1.5,2\}$ (the solid, dashed, and dotted curves). Where this function is positive the inequality is obeyed and where it is negative the inequality is not obeyed. Models with small $\xi$ will never fail to meet the GDSAI criteria everywhere, only across a larger range of radii. In our models \mbox{$\gamma-2\beta>0$ is always true at $r=0$.}}
\end{figure}

In Fig. \ref{fig:hornchangel2plot} we can see that if we raise $\text{L}^2_{cut}$ then it affects less of the system which, in turn, will become increasingly isotropic. Importantly, both the H\'{e}non and radial orbit instabilities are diagnosed by considering the ratio or allowed domain of velocity components. Therefore, if our system is isotropic then it is guaranteed to be stable to both of these effects. This is what we see in Fig. \ref{fig:roicriteriaplot}.

Accordingly, it is the case that the higher $\text{L}^2_{cut}$ is set the greater the proportion of the system that will be isotropic and the larger the volume of the system that will pass those stability criteria. This implies that \emph{the stability of the system is exclusively dependent on the value of the freely tunable parameter $\text{L}^2_{cut}$}. The only limits are the cases of an exactly radial system of infinitesimal density and a completely isotropic system \mbox{\emph{i.e.} $0<\text{L}^2_{cut}\leq Max\left[\text{L}^2_{max}(r)\right]$}.

Thus, one can construct a system using this DF that does not obey the relation of the GDSAI to a specific degree over a specific set of radii by choosing a large enough value for $\text{L}^2_{cut}$. It thus follows that a system could be constructed which produces a small degree of violation over such a small range of radii that it would only be susceptible to the H\'{e}non and radial orbit instabilities to a vanishingly small degree. It is thus true to say that the stability implications become negligible as $\text{L}^2_{cut}\rightarrow Max\left[\text{L}^2_{max}(r)\right]$.

This is particularly true for the Radial Orbit Instability criteria as the susceptibility is averaged over the entire system. Additionally, the H\'{e}non criteria is only necessary, not sufficient, for instability \citep{Merritt1999} and it is thus difficult to claim that small, highly local violations of the criteria represent chronic instability in the model.

This is seen clearly if the GDSAI function is plotted for a selection of our models. In Fig. \ref{fig:hornchangel2plot} we can see the impact of changing $\xi$ on our models. Regions where the function is negative indicates that system is not following the criteria of the GDSAI. In particular, we have shown that the model we have been investigating where $\xi=2$ is stable against radial orbit instability and also fails to follow the same criteria as the GDSAI over a range of radii. In particular, the difference at large radii is significant.

The DF of the model is positive, the majority of the system is isotropic, and the system fails to adhere to any similar relationship to the GDSAI. The only problem with the system is that its stability cannot be rigorously guaranteed. We believe that this proves that the slope-anisotropy inequality cannot be extended to include all non-separable systems in addition to its current areas of success. We obviously cannot speak specifically for each and every non-separable model, but we can demonstrate that DFs of this form with non-trivial values for $\text{L}_{cut}$ do not obey such a relationship. We find that stability criteria are the principle measure of the success of our system and correlate to regions of the system which fail to obey a GDSAI-like relation.

\section{Generalising the model}

We now aim to generalise our DF so as to examine a wider variety of non-separable systems. Additionally we would like to identify a non-separable system that does not follow a GDSAI-like relation whilst also retaining the quality of dynamical stability.

The problems with instability cannot be resolved by using a mono-energy DF. Remember that we established that our systems do not obey anything similar to the GDSAI by excluding high angular momentum orbits to create a density plateau and radial anisotropy. Given that is the case we can generally describe our DF as a mono-energy halo of energy $\text{E}_1$:

\begin{equation}
f(E,L) = \delta(E_1-E) F\left(L^2\right)
\end{equation}
where we require that $F(\text{L}^2)$ decreases as $\text{L}^2$ rises which is how we specify that the model will favour low angular momentum orbits. We can then use this DF to find a general expression for the probability distribution of radial velocities as required for an analysis of the H\'{e}non instability (see \S\ref{sec:Henon}):

\begin{equation}
P(v_r,r) = 2\pi\int \delta\left(E_1-\Phi(r)-\frac{v_r^2}{2}-\frac{v_t^2}{2}\right) F\left(v_t^2 r^2\right) \mathrm{d}\left(\frac{v_t^2}{2}\right)
\end{equation}

This then gives a general solution:

\begin{gather}
P(v_r,r)=F\left(v_t^2 r^2\right) \big| _{\frac{v_t^2}{2}=E_1-\Phi(r)-\frac{v_r^2}{2}}\nonumber\\[5pt]
\quad\quad\quad{}=F\left(\left(2E_1-2\Phi-v_r^2\right)r^2\right)
\end{gather}

Now, we specified that $F(\text{L}^2)$ is a function which decreases as its argument increases. This means that for a fixed radius $r$, $F\left(\left(2E_1-2\Phi-v_r^2\right)r^2\right)$ is an increasing function of $v_r$. This has the unfortunate implication that \mbox{$F(L^2)|_{v_r=0}<F(L^2)|_{v_r>0}$}.

In other words the velocity distribution will always have a trough at $v_r=0$ and two peaks at $v_r=\pm\sqrt{2E_1-2\Phi}$ meaning the H\'{e}non instability is always going to cause problems for models of this kind.

To try and avoid this we must generalise the model further by weakening the condition that it is mono-energy. Our generalised DF is of the form:

\begin{equation}
\label{eqn:newDF}
f(E,L)=\left[H(E-E_1)-H(E-E_2)\right]H\left(L^2_{cut}-L^2\right)g(E)
\end{equation}
where $g(E)$ is a function of energy which we assume, for illustrative purposes, is given by $g(E)=e^{-bE}$. We define $\text{E}_1<\text{E}_2\leq0$ as constant values for energy. With this DF we allow orbits in the system that have angular momentum $\text{L}^2<\text{L}_{cut}^2$, energies of $\text{E}_1<\text{E}<\text{E}_2$ and the function $g(E)$ is left free. We can see that the DF of Eq. \ref{eqn:DF} can be approximated by the special cases where $\text{E}_1 \rightarrow \text{E}_2$ and $g(E)$ is constant.

The corresponding density function for this illustrative model is:

\begin{equation}
\begin{array}{l l}
\label{eqn:illustrative}
\rho=\bigg\{4\pi\sqrt{2}b^{-1.5} \\[10pt]
\quad \quad \quad {\Big[e^{-b\Phi}\left[\Gamma(1.5,b(E_1-\Phi)) - \Gamma(1.5,b(E_2-\Phi))\right]} \\[5pt]
\quad \quad \quad -e^{-b\phi}\left[\Gamma(1.5,b(E_1-\phi)) - \Gamma(1.5,b(E_2-\phi))\right]\Big]\bigg\}
\end{array}
\end{equation}

where $\phi=\Phi+\frac{L^2_{cut}}{2r^2}$, $b$ is a constant, and $\Gamma(a,x)$ is the incomplete Gamma function which is defined as \mbox{$\Gamma(a,x)=\int_x^{\infty}t^{a-1}e^{-t}\,\mathrm{d}t$}.

The corresponding potential is numerically derived and the other system characteristics are computed using the same methods as in \S\ref{sec:DF}. The analytical formulae are not given here as they are prohibitively large and the model is only for illustration.

This generalised DF is non-separable and still demonstrates behaviour that is not in agreement with an extended GDSAI. For example, recall how in the density peak of Fig. \ref{fig:ineqhorndenplot} we had a transition domain where $\gamma\ll0$ while the anisotropy profile was making a transition from strongly radial anisotropy to isotropy as was discussed in detail in \S\ref{sec:result}.

This behaviour remains unchanged in the generalised DF as the angular momentum limit can be set so that it removes \emph{all} particles whose orbits are highly tangential at intermediate radii, making the system underdense and anisotropic. Again, since this cut \emph{only} removes highly tangential orbits it can \emph{only} make the system more radially anisotropic at these radii. Thus the failure to obey a relation similar to the GDSAI is still seen in our generalised system in the region where the density peak is produced as the system moves out of the underdense domain and towards isotropy.

\begin{figure}
\includegraphics[width=84mm]{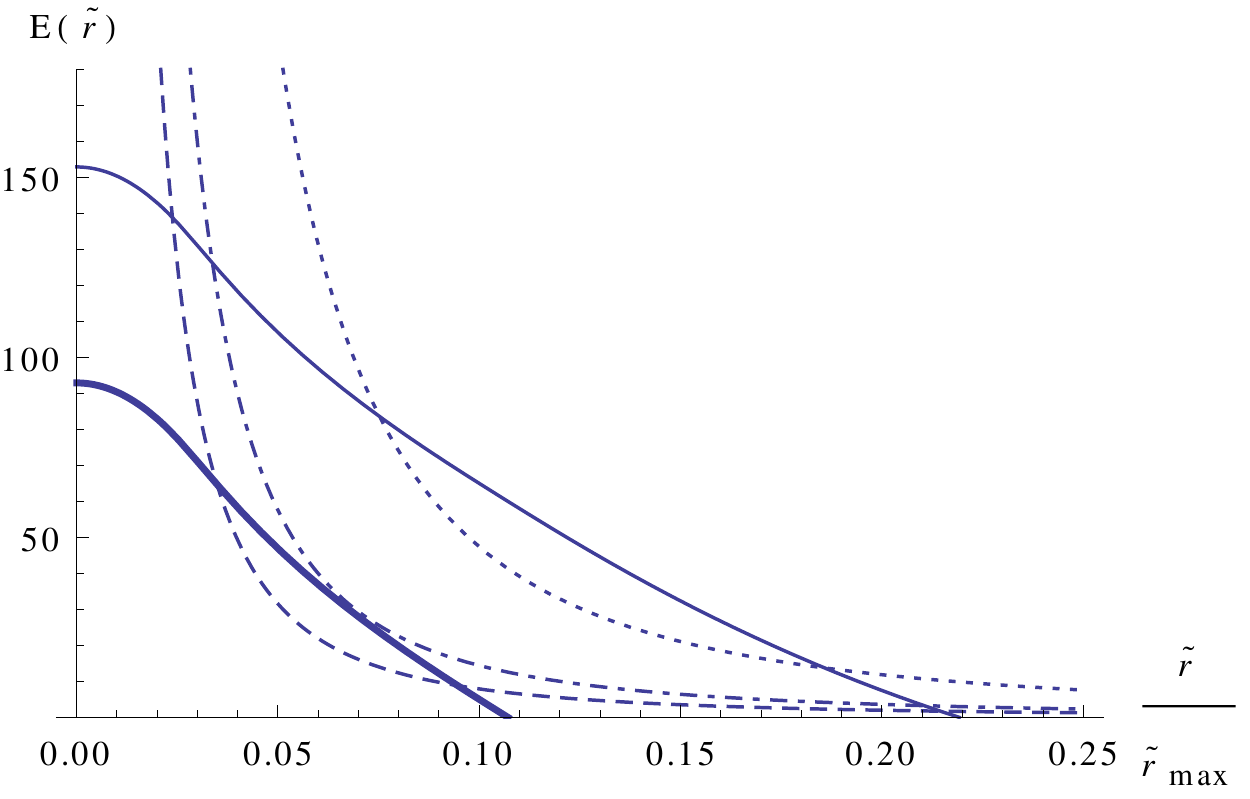}
\caption[Energy functions in the generalised DF]{\label{fig:generalenergy}Plots of important energies for a paricular case of the illustrative model of Eq. \ref{eqn:illustrative} where $\text{E}_1=-70$, $\text{E}_2=-10$, $\text{b}=0.0005$, and $\tilde{\Psi}(\tilde{r}=0)=(4\pi)^2$. The solid and thicker solid curves are $\text{E}_1-\Phi$ and $\text{E}_2-\Phi$. Other curves are $\text{L}_{cut}^2/2r^2$ for different values of $\text{L}_{cut}^2$. The dashed line is for a harsh cut of $\text{L}_{cut}^2=0.012566$. The dot-dashed line is $\text{L}_{cut}^2=0.022934$ which only touches $\text{E}_1-\Phi$. The dotted curve is $\text{L}_{cut}^2=0.075398$ which only excludes a very small amount of orbits.}
\end{figure}

Accordingly, behaviour different to that described by the GDSAI in this region can be caused by simply setting $\text{L}_{cut}^2\leq2\text{R}^2(\text{E}_1-\Phi)$ where R is some chosen intermediate radius at which the angular momentum of a completely tangentially moving particle in the system is maximised. This causes the angular momentum cut to exclude all orbits which are highly tangential around this radius down to the ones of lowest allowed energy as shown in Fig. \ref{fig:generalenergy}. Note that, as demonstrated in Fig. \ref{fig:hornchangel2plot}, the difference from the GDSAI will actually begin at smaller radii due to the gradual flattening of the density profile occurring alongside the rise in anisotropy. However, the exact point at which this takes place will be very model dependent.

\begin{figure}
\includegraphics[width=84mm]{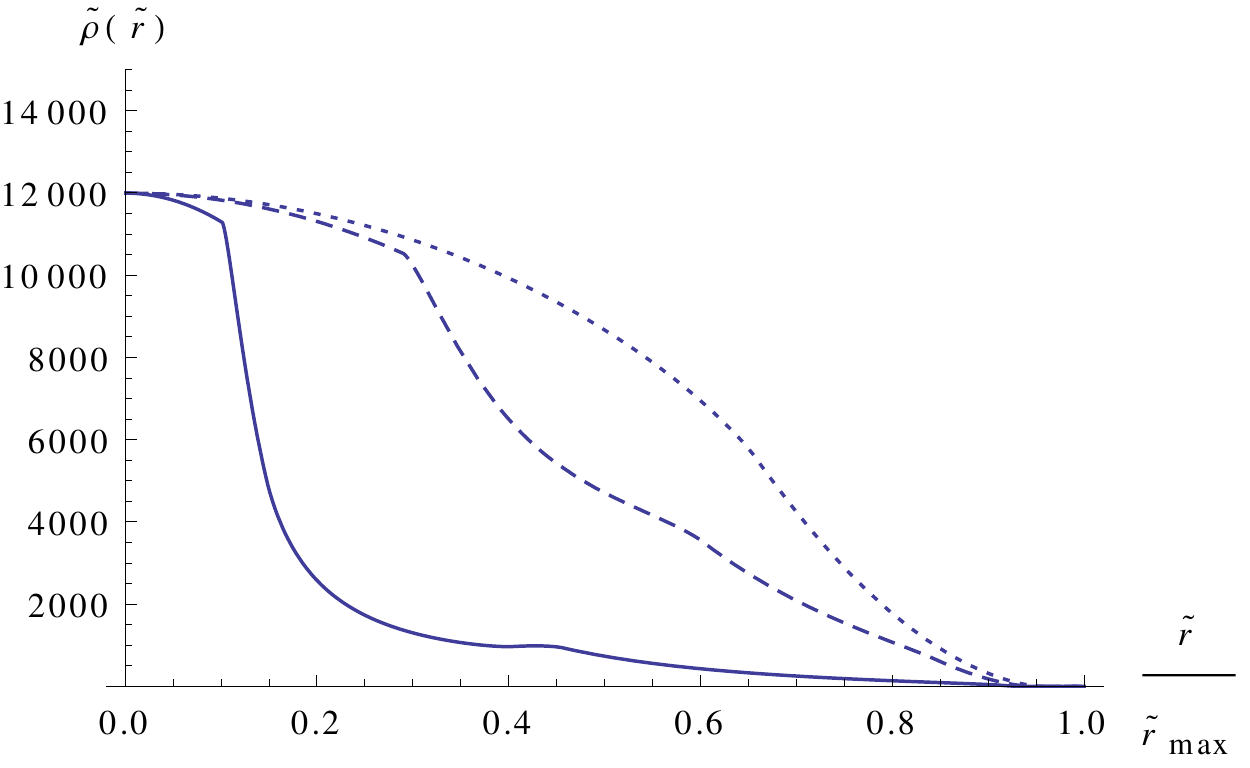}
\caption[Density profiles of our generalised DF]{\label{fig:generaldensity}The density models of the systems in Fig. \ref{fig:generalenergy}. The systems are all $\text{E}_1=-70$, $\text{E}_2=-10$, $\text{b}=0.0005$, and $\tilde{\Psi}(\tilde{r}=0)=(4\pi)^2$ where the solid line is $\text{L}_{cut}^2=0.012566$, the dashed line is $\text{L}_{cut}^2=0.022934$, and the dotted curve is $\text{L}_{cut}^2=0.075398$. Note the bump at intermediate radii rather than the sharp peaks seen in previous models as well as the difficulty in getting a positive gradient.}
\end{figure}

The problem remains that, as discussed in \S\ref{sec:Henon}, removing all highly tangential particles implies that $P(v_r\approx0)=0$ which leads to the sharply double-peaked velocity disitributions that are indicative of the H\'{e}non instability. So, if we set a harsh angular momentum limit then we can remove a large amount of tangential energy and create our peak and anisotropy which will not follow the GDSAI at the cost of stability. Conversely, if we remove the limit entirely then the system becomes isotropic and smooth but fails to produce any interesting behaviour relating to the inequality. However, while these were the only options for the original DF, the new DF allows intermediate cases which can give us some insight as shown in Figs. \ref{fig:generaldensity} and \ref{fig:generalenergy}.

We examine the case where \mbox{$\text{L}_{cut}^2=2r^2(\text{E}_1-\Phi)$} has a single solution compared to the usual two as shown in Fig. \ref{fig:generalenergy}. In this instance the angular momentum limit only excludes orbits down to the orbit of lowest energy which has the highest angular momentum. In other words, if the limit was raised infinitesimally then it would just be possible to have an orbit with kinetic energy of exactly $\text{E}_1-\Phi(R)$ whose apocentre was at a radius R which maximised that orbit's angular momentum.

What we see in Fig. \ref{fig:generaldensity} is that already the model is not excluding enough orbits to force a local $\gamma<0$, which is the feature in our models which always guarantees behaviour that can demonstrate disagreement with a GDSAI-like relation. This is not encouraging as this the earliest case where $P(v_r=0)\neq0$ for all energies which means that that this case is still significantly unstable by the H\'{e}non criteria because $P(v_r=0)\ll P(v_r>0)$. However, it is \emph{already} unlikely to produce behaviour different to the GDSAI due to $\gamma>0$ and an anisotropy that will not be as high as in models with harsher angular momentum cuts.

This is true of every model where \mbox{$2r^2(\text{E}_1-\Phi)\ll \text{L}_{cut}^2<2r^2(\text{E}_2-\Phi)$}. As the number of high angular momentum orbits allowed increases, it follows that $\gamma$ must rise, $\beta$ must tend towards 0, and $P(v_r=0)$ will also rise. However the probability of very low $v_r$ \emph{will still be reduced compared to a system where no orbits are removed}. In other words, the instability will be present to some degree over some range of radii if the angular momentum threshold excludes any orbits at all.

Between these cases we see that the behaviour of our generalised model is comparable to that of our simpler, more specific DF. Depending on the choice of parameters the system can demonstrate behaviour over a range of radii that is in disagreement with a potential extension of the GDSAI, however causing such behaviour decreases the chance of finding particles with low $v_r$. Cutting a large amount of orbits can guarantee this behaviour at the cost of severe instabilities where $P(v_r\approx0)=0$, while weaker cuts causes milder instability where $P(v_r\approx0)\neq0$ and may fail to demonstrate inconsistency with an extended GDSAI.

This line of reasoning leads us to the conclusion that while any suitable combination of parameters can create systems that do not follow a GDSAI-like relationship the mechanism of removing high angular momentum orbits is never going to produce a model that passes the H\'{e}non criteria.

\section{Summary}

We have managed to construct an non-separable, equilibrium system with $\beta_0<1/2$ using a globally positive DF which demonstrates behaviours inconsistent with an application of the GDSAI. The magnitude of the departure from the GDSAI is dependent on the value of the angular momentum threshold $\text{L}^2_{cut}$, which is also the parameter that controls the stability of the model. It is possible to pick values of this parameter where the majority of the system fails to agree with an extension of the GDSAI but is also unstable, or where the failure is highly local and the instability is negligible. This is a significant expansion on previous work proving the efficacy of the GDSAI in separable systems \citep{Ciotti2010,vanhese2011} and is suggestive that the GDSAI may not be applicable to models with non-separable augmented densities.

We conclude this shows that whether or not a non-separable system obeys the GDSAI does not constitute proof of the positivity or otherwise of the system's DF. We do, however, note that there is a non-trivial relationship between disagreement with an extended GDSAI and the stability of the system. We suggest that GDSAI may not imply phase-space consistency in such systems but may be able to make some predictions of model stability.

Exploring generalisations of the simple system have shown that this approach will not be able to yield a system that is stable under the H\'{e}non criteria. Future work will therefore focus on mechanisms beyond the removal of high angular momentum orbits. In conclusion, we feel that while the GDSAI remains a useful guide for non-separable systems, it should not be considered a definitive criterion in discussions of DF positivity in such systems.

\section{Acknowledgements}
The authors would like to thank Steen Hansen of the Dark Cosmology Centre for his helpful discussions and insight. Thanks also to Xufen Wu for her help and input. Thanks to Luca Ciotti and Emmanuel Van Hese for their assisstance in the latter stages of publication.

\bibliographystyle{mn2e}
\bibliography{inequality}

\begin{thebibliography}{}

\bibitem[\protect\citeauthoryear{An \& Evans}{An \& Evans}{2006}]{An2006}
An J.~H.,  Evans N.~W.,  2006, ApJ, 642, 752

\bibitem[\protect\citeauthoryear{Barnes, Hut \& Goodman}{Barnes
  et~al.}{1986}]{BarnesGoodmanHut1986}
Barnes J.,  Hut P.,    Goodman J.,  1986, ApJ, 300, 112

\bibitem[\protect\citeauthoryear{Binney \& Tremaine}{Binney \&
  Tremaine}{2008}]{BinneyTremaine}
Binney J.,  Tremaine S.,  2008, Galactic Dynamics, second edn.
Princeton University Press

\bibitem[\protect\citeauthoryear{Ciotti \& Morganti}{Ciotti \&
  Morganti}{2009}]{Ciotti2009}
Ciotti L.,  Morganti L.,  2009, MNRAS, 393, 179

\bibitem[\protect\citeauthoryear{Ciotti \& Morganti}{Ciotti \&
  Morganti}{2010a}]{Ciotti2010a}
Ciotti L.,  Morganti L.,  2010a, MNRAS, 401, 1091

\bibitem[\protect\citeauthoryear{Ciotti \& Morganti}{Ciotti \&
  Morganti}{2010b}]{Ciotti2010}
Ciotti L.,  Morganti L.,  2010b, MNRAS, 408, 1070

\bibitem[\protect\citeauthoryear{Ciotti \& Morganti}{Ciotti \&
  Morganti}{2010c}]{Ciotti2010b}
Ciotti L.,  Morganti L.,  2010c, in Bertin G.,  {de Luca} F.,  Lodato G.,
  Pozzoli R.,   {Rom{\'e}} M.,  eds, American Institute of Physics Conference
  Series Vol.~1242 of American Institute of Physics Conference Series, {On the
  global density slope-anisotropy inequality}.
pp 300--305

\bibitem[\protect\citeauthoryear{Ciotti \& Pellegrini}{Ciotti \&
  Pellegrini}{1992}]{Ciotti1992}
Ciotti L.,  Pellegrini S.,  1992, MNRAS, 255, 561

\bibitem[\protect\citeauthoryear{Cuddeford}{Cuddeford}{1991}]{Cuddeford1991}
Cuddeford P.,  1991, MNRAS, 253, 414

\bibitem[\protect\citeauthoryear{Dejonghe}{Dejonghe}{1987}]{Dejonghe1987}
Dejonghe H.,  1987, MNRAS, 224, 13

\bibitem[\protect\citeauthoryear{Doremus, Baumann \& Feix}{Doremus
  et~al.}{1973}]{DoremusFeixBaumann1973}
Doremus J.~P.,  Baumann G.,    Feix M.~R.,  1973, A \& A, 29, 401

\bibitem[\protect\citeauthoryear{Doremus, Feix \& Baumann}{Doremus
  et~al.}{1971}]{DoremusFeixBaumann1971}
Doremus J.~P.,  Feix M.~R.,    Baumann G.,  1971, Phys. Rev. Lett., 26, 725

\bibitem[\protect\citeauthoryear{Eddington}{Eddington}{1916}]{Eddington1916}
Eddington A.~S.,  1916, MNRAS, 76, 572

\bibitem[\protect\citeauthoryear{Gillon, Cantus, Doremus \& Baumann}{Gillon
  et~al.}{1976}]{Doremusetal1976}
Gillon D.,  Cantus M.,  Doremus J.~P.,    Baumann G.,  1976, A \& A, 50, 467

\bibitem[\protect\citeauthoryear{Hansen}{Hansen}{2004}]{Hansen2004}
Hansen S.~H.,  2004, MNRAS, 352, L41

\bibitem[\protect\citeauthoryear{Hazewinkel}{Hazewinkel}{1994}]{mathpedia}
Hazewinkel M.,  1994, Encyclopedia of Mathematics (set).
Kluwer

\bibitem[\protect\citeauthoryear{{H\'{e}non}}{{H\'{e}non}}{1973}]{Henon1973}
{H\'{e}non} M.,  1973, A \& A, 24, 229

\bibitem[\protect\citeauthoryear{Hernquist}{Hernquist}{1990}]{Hernquist1990}
Hernquist L.,  1990, ApJ, 356, 359

\bibitem[\protect\citeauthoryear{Merritt}{Merritt}{1999}]{Merritt1999}
Merritt D.,  1999, Pub. Astro. Soci. Pacific, 111, pp. 129

\bibitem[\protect\citeauthoryear{Merritt \& Aguilar}{Merritt \&
  Aguilar}{1985}]{MerrittAguilar1985}
Merritt D.,  Aguilar L.~A.,  1985, MNRAS, 217, 787

\bibitem[\protect\citeauthoryear{Polyachenko, Polyachenko \&
  Shukhman}{Polyachenko et~al.}{2013}]{Polyachenko2013}
Polyachenko E.~V.,  Polyachenko V.~L.,    Shukhman I.~G.,  2013, MNRAS, 434,
  3208

\bibitem[\protect\citeauthoryear{Taylor \& Navarro}{Taylor \&
  Navarro}{2001}]{Taylor2001}
Taylor J.~E.,  Navarro J.~F.,  2001, ApJ, 563, 483

\bibitem[\protect\citeauthoryear{{Van Hese}, Baes \& Dejonghe}{{Van Hese}
  et~al.}{2011}]{vanhese2011}
{Van Hese} E.,  Baes M.,    Dejonghe H.,  2011, ApJ, 726, 80

\bibitem[\protect\citeauthoryear{Zhao}{Zhao}{1996}]{Zhao1996}
Zhao H.,  1996, MNRAS, 278, 488

\end{thebibliography}


\appendix

\section{Substitutions for deriving the anisotropy}

For clarity we solve each integral individually as they require substitutions to be easily soluble. We consider the numerator first and make a substitution of $Y=L^2/L^2_{cut}$ to make the problem dimensionless:

\begin{gather}
\mathbb{I}_1=\int^{L^2_{cut}}_0\frac{L^2}{r^2\sqrt{E_0-\Phi(r)-\frac{L^2}{2r^2}}}\,\mathrm{d}L^2\nonumber\\[5pt]
\quad\quad\quad\quad\quad{}=\frac{L^4_{cut}\sqrt{2}}{r\sqrt{L^2_{cut}}}\int^1_0\frac{Y}{\sqrt{\frac{2r^2}{L^2_{cut}}\left[E_0-\Phi(r)\right]-Y}}\,\mathrm{d}Y
\end{gather}

Similarly for the denominator we make the same substitution:

\begin{gather}
\mathbb{I}_2=2\sqrt{2}\int^{L^2_{cut}}_0\sqrt{E_0-\Phi(r)-\frac{L^2}{2r^2}}\,\mathrm{d}L^2\nonumber\\[5pt]
\quad\quad\quad\quad{}=\frac{2 L^3_{cut}}{r}\int^1_0\sqrt{\frac{2r^2}{L^2_{cut}}\left[E_0-\Phi(r)\right]-Y}\,\mathrm{d}Y
\end{gather}

These integrals are simpler to evaluate and yield the results of \S3.2

\bsp

\label{lastpage}

\end{document}